\documentclass[aip,reprint,graphicx]{revtex4-1}

\usepackage{graphicx}
\usepackage{amsmath}
\usepackage{amssymb}

\DeclareMathOperator{\sgn}{sgn}

\newcommand\rRi{r_{\text{R},i}}
\newcommand\rRf{r_{\text{R},f}}
\newcommand\rT{r_\text{T}}

\newcommand\NL{N_\text{L}}
\newcommand\NF{N_\text{F}}
\newcommand\NI{N_\text{I}}

\newcommand\NRi{N_{\text{R},i}}

\newcommand\NT{N_\text{T}}
\newcommand\NTot{N_\text{Tot}}

\newcommand\Nesc{N_\text{e}}

\newcommand\LT{L_\text{T}}
\newcommand\Lp{L_\text{p}}

\newcommand\rp{r_\text{p}}
\newcommand\rw{R_\text{w}}
\newcommand\rL{r_\text{L}}
\newcommand\omegac{\omega_\text{c}}
\newcommand\fc{f_\text{c}}
\newcommand\omegar{\omega_\text{r}}
\newcommand\fr{f_\text{r}}
\newcommand\omegas{\omega_\text{s}}
\newcommand\baromegas{{\bar \omega}_\text{s}}
\newcommand\baromegar{{\bar \omega}_\text{r}}
\newcommand\omegap{\omega_\text{p}}
\newcommand\omegaz{\omega_\text{z}}
\newcommand\fz{f_\text{z}}

\newcommand\kB{k_\text{B}}

\newcommand\EB{E_{\text{B}}}

\newcommand\me{m_\text{e}}

\newcommand\sigmaL{{\sigma_\text{L}}}
\newcommand\AL{{A_\text{L}}}
\newcommand\FP{{F_\text{P}}}

\begin{document}

\title{Electron Cyclotron Resonance (ECR) Magnetometry with a Plasma Reservoir} 

\author{E. D. Hunter}
\affiliation{Department of Physics, University of California, Berkeley, California, 94720 USA}
\author{A. Christensen}
\affiliation{Department of Physics, University of California, Berkeley, California, 94720 USA}
\author{J. Fajans}
\email{joel@physics.berkeley.edu}
\affiliation{Department of Physics, University of California, Berkeley, California, 94720 USA}
\author{T. Friesen}
\affiliation{University of Calgary, Calgary, Alberta T2N 1N4 Canada}
\author{E. Kur}
\affiliation{Department of Physics, University of California, Berkeley, California, 94720 USA}
\author{J. S. Wurtele}
\affiliation{Department of Physics, University of California, Berkeley, California, 94720 USA}

\date{\today}

\begin{abstract}
The local magnetic field in a Penning-Malmberg trap is found by measuring the temperatures that result when electron plasmas are illuminated by microwave pulses. Multiple heating resonances are observed as the pulse frequencies are swept.  The many resonances are due to electron bounce and plasma rotation sidebands. The heating peak corresponding to the cyclotron frequency resonance is identified to determine the magnetic field. A new method for quickly preparing low density electron plasmas for destructive temperature measurements enables a rapid and automated scan of microwave frequencies. This technique can determine the magnetic field to high precision, obtaining an absolute accuracy better than $1\,\mathrm{ppm}$, and a relative precision of $26\,\mathrm{ppb}$.  One important application is in situ magnetometry for antihydrogen-based tests of charge-parity-time symmetry and of the weak equivalence principle.
\end{abstract}

\maketitle 

\section{Introduction}
Measuring the magnetic field magnitude in a Penning-Malmberg trap\cite{malm:82,davi:90} is of direct interest to many nonneutral plasma and neutral-trap experiments, particularly to the fundamental physics experiments being conducted by the ALPHA (Antihydrogen Laser Physics Apparatus) collaboration\cite{amol:14b} at CERN's Antiproton Decelerator (AD). Accurate measurements of the magnetic field are critical to precision measurements of the atomic spectra,\cite{ahma:17b,ahma:18a,ahma:18b} and the gravitational acceleration of antihydrogen.\cite{amol:13,zhmo:13,hami:14} These measurements constitute important tests of charge-parity-time (CPT) symmetry and of Einstein's weak equivalence principle.

After a brief introduction to the experiment (Sec.~\ref{sec:Exp}), we describe a method for quickly preparing a sequence of target pure-electron plasmas from a large plasma reservoir (Sec.~\ref{sec:Res}). We then heat each target plasma with a microwave pulse, sweeping the microwave frequency $F$ between plasmas, and measure the resulting plasma temperature $T$ (Sec.~\ref{sec:uHeat}). The reservoir technique makes it possible to perform a complete $T$ versus $F$ scan in about a minute.

We observe a sequence of peaks separated by the axial bounce frequency of the electrons in their trapping potential, and explain their origin for the purpose of identifying the peak corresponding to the electron cyclotron resonance (ECR) frequency,\cite{onei:80a,davi:90,goul:95,dubi:13} $\omegac=2\pi\fc=eB/\me$. Here $\mathbf{B}=B\hat{z}$ is the magnetic field, and $-e$ and $\me$ are the charge and mass of the electron respectively (Sec.~\ref{sec:Bounce}). We then discuss the presence of subpeaks separated by the plasma rotation frequency, and identify the subpeak corresponding to $\omegac$, thus completing our magnetometry measurement. We propose an explanation for the qualitative differences between our observed subpeaks and those predicted by Davidson\cite{davi:90} and Gould\cite{goul:95} and experimentally verified in a variety of NNP systems\cite{goul:92,sari:95,affo:15} (Sec.~\ref{sec:Rot}). Next, we employ these methods to perform precision magnetometry in our electron plasma trap (Sec.~\ref{sec:Mag}). We have used this method in the neighborhood of $B=0.16 \,\mathrm{T}$ and $0.7 \,\mathrm{T}$ to measure the resonant $f$ to an accuracy of a few kHz, corresponding to a $B$ accuracy of better than $1 \,\mathrm{ppm}$.

We discuss plasma expansion during reservoir operations in Appendix~\ref{App:ResExpansion}, and counter-rotating modes in Appendix~\ref{App:Negl}. Next, we discuss systematic limitations to ECR magnetometry in Appendix~\ref{App:Errors}, followed by a description of our fitting methods in Appendix~\ref{App:Fit}.  We conclude with a discussion of alternate magnetometry techniques in Appendix~\ref{App:AltTech}.

\section{Experiment}
\label{sec:Exp}
\begin{figure*}
\centering
\includegraphics[scale=.11]{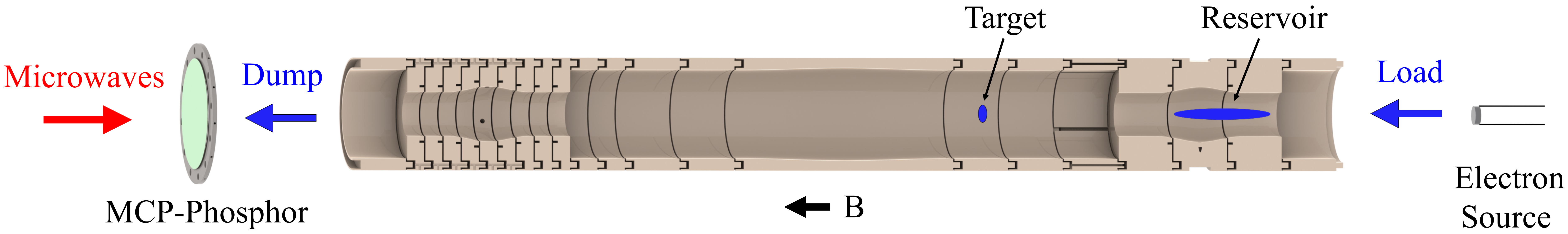}
\caption{The Penning-Malmberg trap. Electrons emitted by the barium oxide cathode of the electron source (far right) are accumulated in the rightmost cavity to form the plasma reservoir. Target plasmas are individually extracted from the reservoir and moved to a heating well. Microwaves enter through the MCP/Phosphor screen (far left) and heat the target plasmas. Target plasma temperatures are measured by reducing the downstream confinement potential and recording the arrival rate at the MCP.  The target plasmas are confined in an electrode of radius $\rw=2\,\mathrm{cm}$ and length $2.54\,\mathrm{cm}$.}
\label{stack}
\end{figure*}

We use a Penning-Malmberg trap (see Fig.~\ref{stack}) to confine our pure-electron plasmas.  Such traps use an axial magnetic field, in our case from a superconducting solenoid, for radial confinement. An electrostatic potential well, formed by a set of individually biasable coaxial cylinders, provides axial confinement.  The biases can be manipulated to move the axial location of the electrostatic well, and hence, the plasma position. The trap is loaded with electrons from an upstream hot-cathode electron source. The entire trap is cooled by attachment to a $4\,\mathrm{K}$ coldhead, which ensures that the electrons are confined under ultra high vacuum (UHV) conditions when the electron source is off.

Destructive measurements of the plasma shape and temperature can be performed by reducing the confinement barrier downstream of the plasma, thereby releasing the plasma towards a microchannel plate (MCP) detector. When electrons strike the MCP, the MCP produces a charge cascade which hits the phosphor screen mounted directly behind the MCP. A CCD camera focused on the phosphor screen is used to image the light that results.\cite{peur:93a} These images are a measure of the $z$-integrated charge density of the plasma. A fitting algorithm\cite{evan:16a} is employed to obtain the  plasma radius.

We measure plasma temperatures by recording the time history of the MCP/Phosphor light, measured with a silicon photomultiplier (SiPM),\cite{hunt:20a} when the downstream potential barrier $\EB=e(V_0-vt)$ is slowly lowered at linear rate $v$ from its initial value $V_0$.  As $\EB$ decreases, the most energetic plasma electrons, those furthest out in thermal distribution, escape first. We assume that the thermal distribution is Maxwellian due to collisions.

\begin{figure}
\centering
\includegraphics[scale=.4]{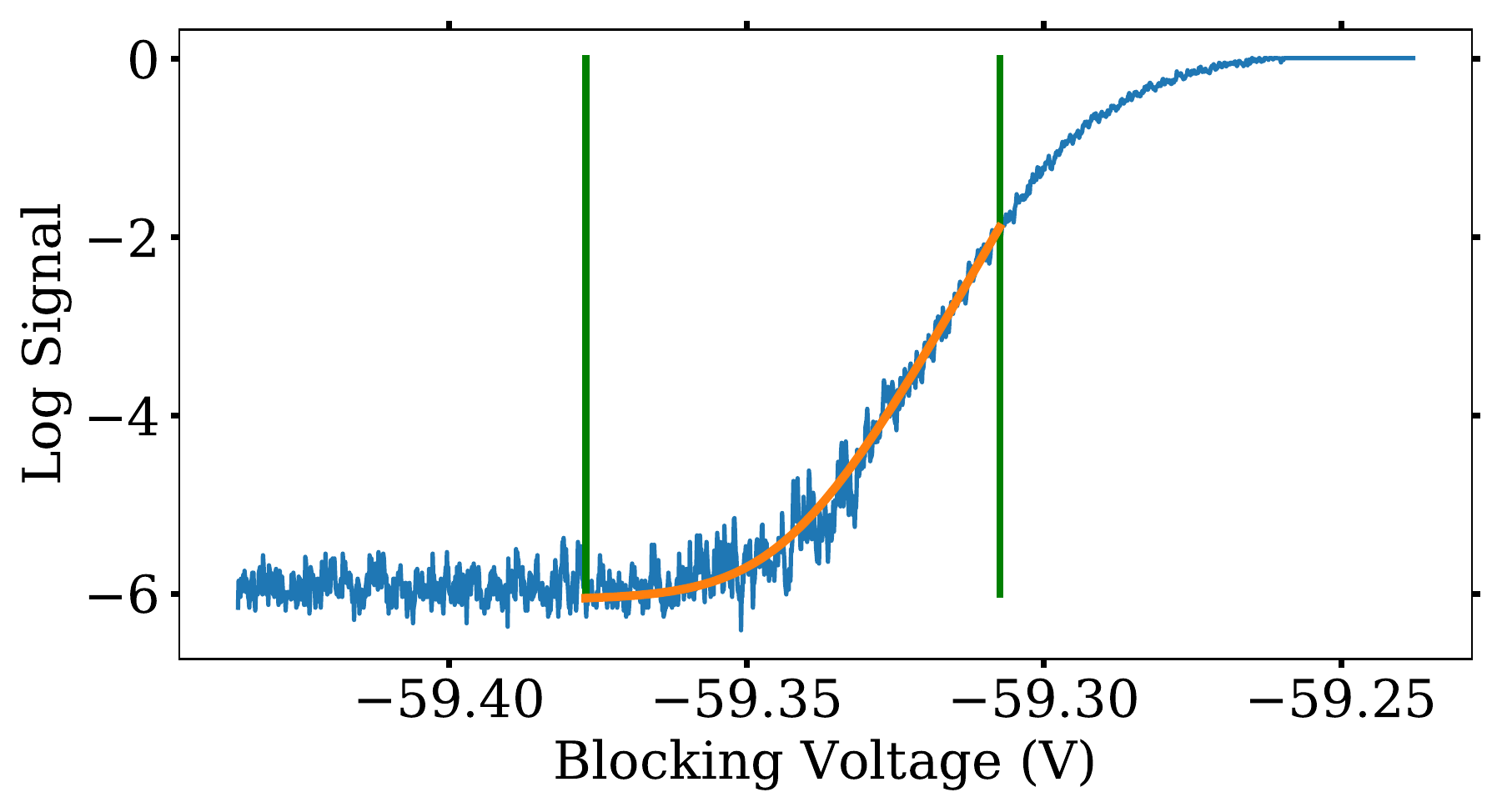}
\caption{A characteristic temperature fit; the Blocking Voltage is shifted from $\EB/e=0$ . The fit region is between the two vertical green lines.  The signal increases linearly for several decades on a log scale.  The temperature $T$ was found to be $99\,\mathrm{K}$ for this data.}
\label{temp}
\end{figure}

Initially, the amount of escaped charge $\Nesc$ is exponential in time $t$ with a rate inversely proportional to the temperature $T$:\cite{eggl:92}
\begin{equation}
\label{eq:temp}
\frac{d\Nesc}{dt}\propto \exp{[-\EB(t)/\kB T]}
\end{equation}
The SiPM provides single-electron resolution, so we can measure temperatures for very low particle number [$\NT\approx\mathcal{O}(10^3)$] plasmas. Figure~\ref{temp} shows a characteristic temperature fit.  The fit is found using the Levenberg-Marquardt
algorithm using a slight generalization of Eq.~(\ref{eq:temp}), $A+B\exp{[-\EB(t)/\kB T]}$, where $A$, $B$ and $T$ are the fitting parameters.  In a slight deviation from general practice, we fit on $\Nesc$ itself, not the more commonly used $\ln{\Nesc}$, because the noise is not proportional to the signal strength. The fitting region is automatically optimized to be between the noise floor at large $\EB$, and where deviations from Eq.~(\ref{eq:temp}) become significant at small $\EB$;\cite{evan:16a}  these deviations develop as significant charge escapes\cite{eggl:92} and as plasma instabilities set in.

We inject microwaves at the MCP end of the trap through a horn attached to an HP~8673d signal generator.
The plasma temperature is measured after injecting the microwaves for a period of $100\,\mathrm{ms}$. Numerous ECR heating peaks are observed during a typical frequency scan. As discussed later, we associate the peaks with the bounce and rotation frequency sidebands of the cyclotron frequency. By varying the confining electrode potentials we identify the peak corresponding to the fundamental cyclotron frequency. We can then determine the magnetic field magnitude through inversion of the cyclotron frequency formula, $B = (2\pi\me/e)\fc$.

\section{Reservoir}
\label{sec:Res}
ECR magnetometry has been employed in a Penning-Malmberg trap before,\cite{amol:14b,frie:14} primarily using a non-destructive, plasma modes-based temperature measurement technique.\cite{dubi:91,tink:94}  However, in many, perhaps all Penning-Malmberg traps, the modes-based diagnostic only works with a target plasma with a large number of electrons ($\NT\approx 2\times 10^6$ in our experiment).  The sidebands in such a large plasma may be difficult to separate.  Moreover, such plasmas are physically large, and may span a broad range of cyclotron frequencies if the magnetic field is inhomogeneous.  Thus, accurate, local, magnet field measurements require a small plasma, for which only destructive temperature measurements can be made.

The most straightforward method to generate the new target plasmas required for each destructive measurement is to load each plasma directly from the electron source.  However this requires turning the source on (a few second process to warm up), or leaving it on and hot, which degrades the trap vacuum and heats the cryogenic trap.   After capturing each plasma from the source,  its parameters must be tailored appropriately.  The time required to individually generate all the target plasmas necessary for a frequency scan could be hours. Even if the time is available, drifting magnetic fields may limit the measurement resolution.

Drawing the target plasmas from a plasma reservoir avoids many of these time-consuming steps.  For each frequency scan, we use the electron source only once to prepare a plasma reservoir.  We can then extract over one hundred target plasmas from the reservoir at a rate as fast as ten target plasmas per second.

Our reservoir typically contains $20\mbox{--}30\times 10^6$ electrons, from which we withdraw and, after further processing, capture target plasmas with as few as $\NT=1500$ electrons.  The initial steps in our method resemble those used by Danielson et al.\ to extract small diameter beams from positron plasmas,\cite{dani:07} but they do not then capture the particles in their beam into a plasma.  The BASE collaboration employs a much smaller (typically $\sim 100$ antiprotons) reservoir from which they repetitively draw single antiprotons.\cite{smor:15}

\subsection{Reservoir Plasma Preparation}

Before drawing target plasmas, we must stabilize the number of particles, the temperature, and the density of our reservoir plasmas. To accomplish this, we use a technique called strong-drive regime evaporative cooling (SDREVC).\cite{ahma:18} This technique involves applying a strong drive (SDR) rotating electric field to fix the plasma rotation frequency and density,\cite{dani:05} while simultaneously performing forced evaporative cooling\cite{andr:10} (EVC) to control the plasma space charge.  EVC, with assistance from cyclotron cooling, also keeps the plasma temperature below $1000\,\mathrm{K}$, a prerequisite for SDREVC.  (We typically use magnetic fields of approximately $0.7 \,\mathrm{T}$, where cyclotron cooling, which scales as $|B|^2$, is not as effective as it was in the $1\mbox{--}3\,\mathrm{T}$ fields of Ref.~\onlinecite{ahma:18}.)

\begin{figure}
\centering
\includegraphics[scale=.4]{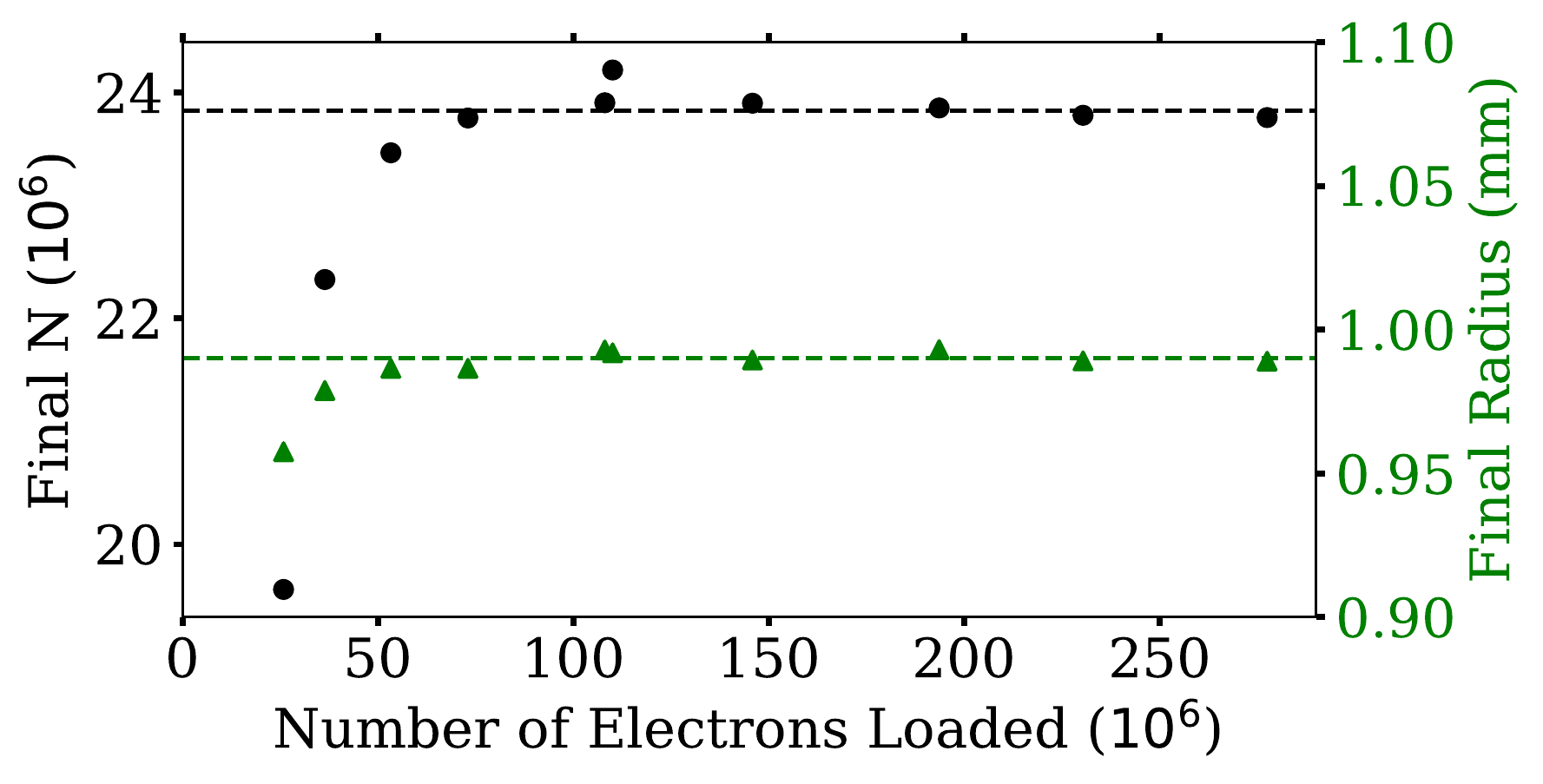}
\caption{Plasma parameters post SDREVC as a function of the number of electrons initially loaded $\NL$. For $\NL>75 \times 10^6$, SDREVC reproducibly reduces the reservoir to a fixed particle number $\NF$ (black dots), radius (green triangles), and temperature (not shown). For $\NL<40 \times 10^6$, the preliminary evaporation step does not remove any particles, and the plasma is not cold when SDREVC begins: hence, the imperfect stabilization. Dotted lines show the limiting values of the number and radius.}
\label{sdrevc}
\end{figure}

We begin our reservoir preparation by loading a plasma of $\NL\approx 200\times 10^6$ electrons, and aggressively evaporate it down to some tens of millions of electrons to obtain a much reduced temperature of $T \approx 500 \,\mathrm{K}$.  We then perform SDREVC, tuning the sequence of potentials to maintain $T$ as low as possible. In Fig.~\ref{sdrevc} we show the post SDREVC plasma parameters as a function of the number of initially loaded electrons for a sequence optimized at $B=0.7 \,\mathrm{T}$; the final number of electrons in the reservoir is $\NF=24\times 10^6$.  We note that this sequence still works at the $5$\% level ($\Delta\NF/\NF$), for a similar range of initial $\NL$, at $B=0.16 \,\mathrm{T}$. The cooling time in this field exceeds $100\,\mathrm{s}$, so the cooling here must be almost entirely evaporative.

\subsection{Extracting Target Plasmas}
\label{sec:extr}
\begin{figure}
\centering
\includegraphics[scale=.4]{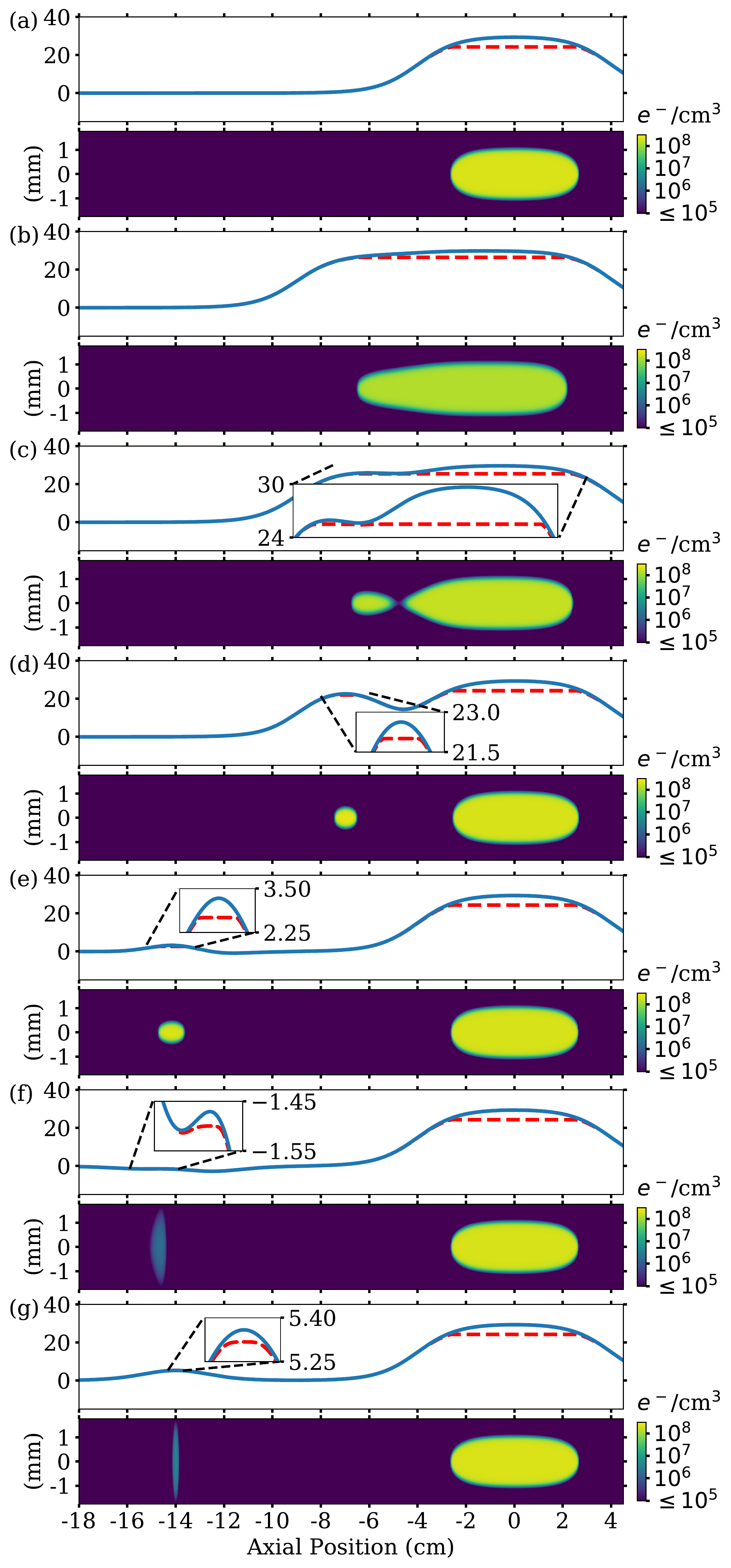}
\caption{The reservoir extraction process.  Each step (a---g) (see Sec.~\ref{sec:extr}) is illustrated by a set of two plots.  The on-axis potential is shown in volts in the upper plot in each set. The solid blue lines show the vacuum potential while the dashed red lines show the total potential including the plasma charge. The insets show vertical expansions for the potential.  The limited-radius trap cross section is shown in the lower plot in each set. The plasma density is plotted using color as a function of axial (horizontal) and transverse (vertical) position. }
\label{fig:reservoir_extraction}
\end{figure}

Figure~\ref{fig:reservoir_extraction}, shows the seven plasma manipulation steps we use to extract a target plasma from the reservoir. The graphs in this figure were generated using a grid-based numerical solver\cite{pras:79,peur:90,spen:93} which determines the plasma density in thermal and rotational equilibrium in an infinite cylindrical trap with given electrode voltages and lengths.

The seven manipulation steps are as follows:
\begin{enumerate}
\item[a.] The prepared reservoir plasma is shown in its resting state.
\item[b.] The reservoir is elongated so that it extends across three electrodes. The leftmost electrode will ultimately confine the target plasma, the rightmost electrode will ultimately confine the reservoir plasma, and the center electrode will be used to separate the two plasmas. In this example, the reservoir electrode is set to $+30\,\mathrm{V}$, while the other two electrodes are set to $+28\,\mathrm{V}$.  This diminishes the radius of the plasma under these electrodes.
\item[c.] The voltage on the center electrode is decreased to cut the plasma.  The image shown occurs just after the plasma is split.
\item[d.] The voltage on the center electrode is further decreased until the plasmas are fully separated. We used a linear change in voltage to progress from (b) to (d).
\item[e.] The electrode potentials are ``rolled'' to move the target plasma  a safe distance from the reservoir so that the reservoir does not affect the electric fields felt by the target plasma.
\item[f.] The target plasma is evaporatively cooled to control its temperature and reduce the number of plasma particles to $\NT=1.5\times 10^3$ to $3\times 10^4$. The target plasma radius increases as described in Appendix~\ref{App:ResExpansion}.\cite{andr:10}  Because the plasma has so few particles, this is a delicate step to perform properly, for which we need accurate models of the vacuum potential.
\item[g.] The target plasma is put in a deeper potential well, and it is ready to receive microwaves.
\end{enumerate}
Typically, we extract $60\mbox{--}120$ target plasmas from each reservoir plasma by cycling these steps. As charge is extracted, the reservoir's self-consistent potential (the vacuum and space charge potentials) increases. Therefore, the ``target electrode bias voltage,'' $+28\,\mathrm{V}$ on the left electrode in step b, must be increased for each successive extraction.

\begin{figure}
\centering
\includegraphics[scale=.4]{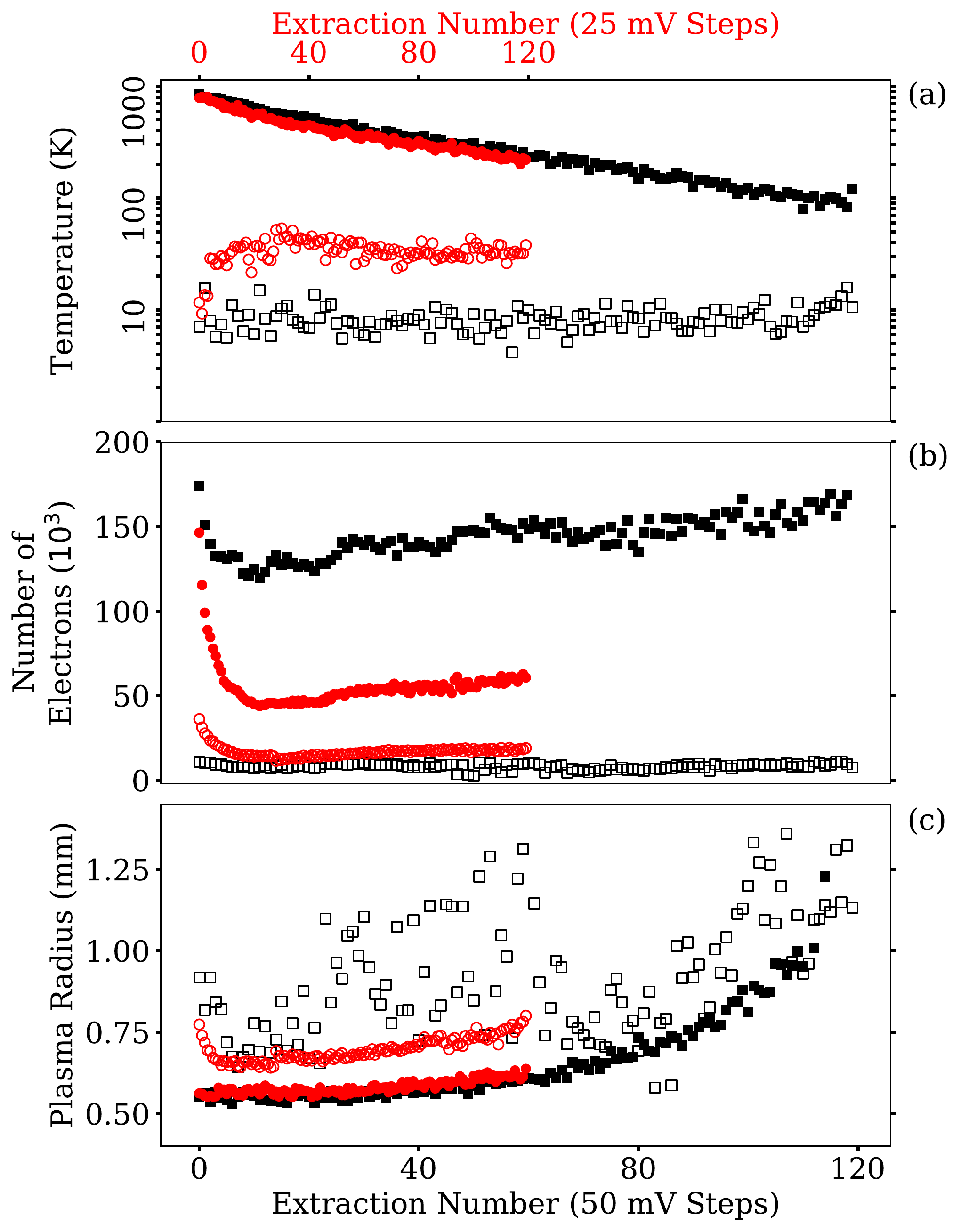}
\caption{The temperature, charge, and radius of the target plasmas as a function of the extraction number.  The target electrode bias voltage (defined in Sec.~\ref{sec:extr}) is increased in steps of $25\, \mathrm{mV}$ (red circles, top axis) and $50\, \mathrm{mV}$ (black squares, bottom axis) for subsequent extractions. The horizontal spacing between the extractions is different for the $25\, \mathrm{mV}$ and $50\, \mathrm{mV}$ datasets in order to place extractions of equal target bias voltage at the same horizontal position. The plasma parameters are shown before (solid symbols) and after (hollow symbols) the target plasmas are evaporatively cooled (EVC). In the post-EVC data for the $50\,\mathrm{mV}$ steps, the target plasma density is low and the images on the phosphor screen are correspondingly dim, leading to large extraction-to-extraction uncertainty in the fitted plasma radius.}
\label{Fig:TargetPlasmas}
\end{figure}

Figure~\ref{Fig:TargetPlasmas} plots the measured  temperature, charge, and radius of target plasmas for a sequence of $120$ extractions. The figure shows the effect of choosing $25$ and $50\,\mathrm{mV}$ target bias voltage increment steps; the larger steps initializes the target plasma with more electrons before EVC (step f above). The reservoir is evaporatively cooled by the extraction of target plasmas, so the later-extracted target plasmas are correspondingly colder before EVC. After EVC, the target plasma temperatures are much lower and largely independent of the extraction number. Since the post EVC  potentials are the same for both the $25$ and $50\,\mathrm{mV}$ steps, a larger fraction of the particles are lost during EVC for the $50\,\mathrm{mV}$ steps, resulting in lower plasma temperatures and densities, and larger plasma radii.

As can be seen in Fig.~\ref{Fig:TargetPlasmas}, the plasma parameters are reasonably stable after an initial transient.  We discard the first few extractions; the variations in the subsequent extractions have no noticeable effects on our magnetometry measurements.

In the precision magnetometry results presented below, numerical calculations yield target plasma lengths of $\Lp=0.18$ to $0.10\,\mathrm{mm}$ and densities of $n=2.3\times10^7$ to $5.6\times10^7\,\text{cm}^{-3}$ for plasmas with $\NT=1500$ electrons, radii of $\rp=1\,\mathrm{mm}$, and confined in potentials corresponding to bounce frequencies of $30$ to $55\,\mathrm{MHz}$.  At $10\,\mathrm{K}$, the Debye lengths of these plasmas are $0.05$ to $0.03\,\mathrm{mm}$, and their plasma parameters are of order unity.  Thus, the target ``plasmas'' are insufficiently dense to be solidly in the plasma regime, and will begin to lose some of their collective properties.  For example, the plasmas will not entirely flatten the on-axis potential.  For simplicity, we will nonetheless refer to these charge ensembles as plasmas.

\section{Microwave Heating}
\label{sec:uHeat}
In the presence of an axially-propagating electromagnetic wave $\mathbf{E}(\mathbf{r},t)$, the equation of motion of a single electron in a magnetic field $\mathbf{B}$ is:
\begin{equation}
\label{eq:EoM}
\ddot{\mathbf{r}}=\frac{-e}{\me}\left(\dot{\mathbf{r}}\times\mathbf{B}
                  +k_2\left(z\hat{z}-\frac{1}{2}(x\hat{x}+y\hat{y})\right)+\mathbf{E}\right),
\end{equation}
where we have assumed an approximate, harmonic trapping potential $V=-\frac{1}{2}k_2(z^2-\frac{1}{2}(x^2+y^2))$, ignored the plasma self-field, and also ignored any non-transverse electric and all magnetic components of the electromagnetic wave.

Equation~(\ref{eq:EoM}) trivially decomposes into parallel and transverse equations. The parallel equation is solved by $z=z_0\sin(\omegaz t)$, where the angular bounce frequency is defined by $\omegaz=2\pi\fz=\sqrt{ek_2/\me}$. We can simplify the transverse equation by adopting the notation $\mathbf{X}=X_-(\hat{x}-i\hat{y})\exp(i\omega t)$ for the electron position, and assuming that the applied microwave electric field consists of circularly-polarized plane waves,
\begin{equation}
\label{eq:Eoriginal}
\mathbf{E}=\left[E_-(\hat{x}-i\hat{y})+E_+(\hat{x}+i\hat{y})\right]\exp(i\omega t),
\end{equation}
where we temporarily ignore any $z$ dependencies and the nonresonant $E_+$ term, keeping only the resonant $E_-$ term, which has the same helicity as $\mathbf{X}$.  Then, Eq.~(\ref{eq:EoM}) becomes
\begin{equation}
\label{comp_eom}
-\omega^2{X_-}+\omega\omegac{X_-}-\frac{\omegaz^2}{2}X_-=\frac{-e}{\me}E_-.
\end{equation}
The homogenous solutions of this equation show that an undriven electron executes a fast cyclotron-like motion $\omegac'$ and a slow drift rotation, sometimes called the magnetron rotation, $\omegar=2\pi\fr$.  The well-known frequencies of these motions are\cite{byrn:65,jeff:83,sari:95}
\begin{flalign}
\label{Eq:cyclotron}
\omegac' &=\frac{\omegac}{2}\left[1+\left(1-\frac{2\omegaz^2}{\omegac^2}\right)^{1/2}\right]= \omegac-\omegar, \\
\label{Eq:rotation}
\omegar &=\frac{\omegac}{2}\left[1-\left(1-\frac{2\omegaz^2}{\omegac^2}\right)^{1/2}\right]\approx \frac{\omegaz^2}{2\,\omegac}.
\end{flalign}

We now make a small modification to Eq.~(\ref{eq:EoM}), introducing a damping term $\dot{\mathbf{r}}/\tau$, with a decoherence time $\tau\gg 1/\omegac$, to include the effects of collisions with other electrons, and, possibly, background gas. Equation~(\ref{comp_eom}) then becomes
\begin{equation}
\label{eq:DampedEOM}
-\omega^2{X_-}+\omega\omegac{X_-}-\frac{\omegaz^2}{2}X_-+\frac{i\omega}{\tau}{X_-}=\frac{-e}{\me}E_-.
\end{equation}
This change introduces an exponential decay with timescale $\tau$ to the cyclotron motion at $\omega_c'$.  It also introduces a decay of the rotational motion at $\omegar$ that is roughly $2\omega_c/\omegar\sim 10^6$ times slower that the decay of the cyclotron motion. This decay is unphysical and is an artifact of the crude way that collisions were introduced.

To find the particular solutions of Eq.~(\ref{eq:DampedEOM}), we regroup yielding
\begin{equation}
\left(i\omega-i\omegac'+\frac{1}{\tau}\right)i\omega X_-=\frac{-e}{\me}E_-,
\label{eq:fourtrans}
\end{equation}
where we have assumed that the drive frequency is sufficiently close to the cyclotron frequency that we can approximate $\omegac-\omegaz^2/(2\omega)$ as $\omegac'$. Then, the microwave power $P$ absorbed by an electron is $e\operatorname{Re}(\mathbf{E}) \cdot \operatorname{Re}(\dot{\mathbf{X}})$, where the velocity $\dot{\mathbf{X}}= i\omega\mathbf{X}$ can be found from the solution of Eq.~(\ref{eq:fourtrans}).  Thus,
\begin{equation}
P(\omega)=\frac{e^2\tau}{\me}\frac{|E_-|^2}{1+[\tau(\omegac'-\omega)]^2}.
\label{eq:power}
\end{equation}
Eq.~(\ref{eq:power}) indicates that the linewidth of the heating peak will be set by the decoherence time; note that our microwave illumination time is sufficiently long that it does not affect the linewidth.

In some circumstances, the linewidth can instead be dominated by magnetic field inhomogeneities, which we have not modeled in these equations. The very short target plasmas generated by the reservoir technique mitigate this effect as they sample only a very small region of the inhomogeneous field.

We launch a linearly polarized wave into our experimental system, not a circularly polarized wave as in Eq.~(\ref{eq:Eoriginal}).  After injection, the wave propagates in a highly overmoded structure with many obstacles and we do not maintain control of its polarization or mode structure.  Consequently, we do not know what fraction of the $10\,\mathrm{dBm}$ injected microwaves reaches the target electrons. While Eq.~(\ref{eq:power}) gives us a useful qualitative picture of the plasma heating, we do not use it to relate the lineshapes of our observed heating peaks to the physical parameters of our system, nor do we use it to predict the peak amplitudes.

\section{Bounce Frequency Sidebands}
\label{sec:Bounce}
\begin{figure}
\centering
\includegraphics[scale=.4]{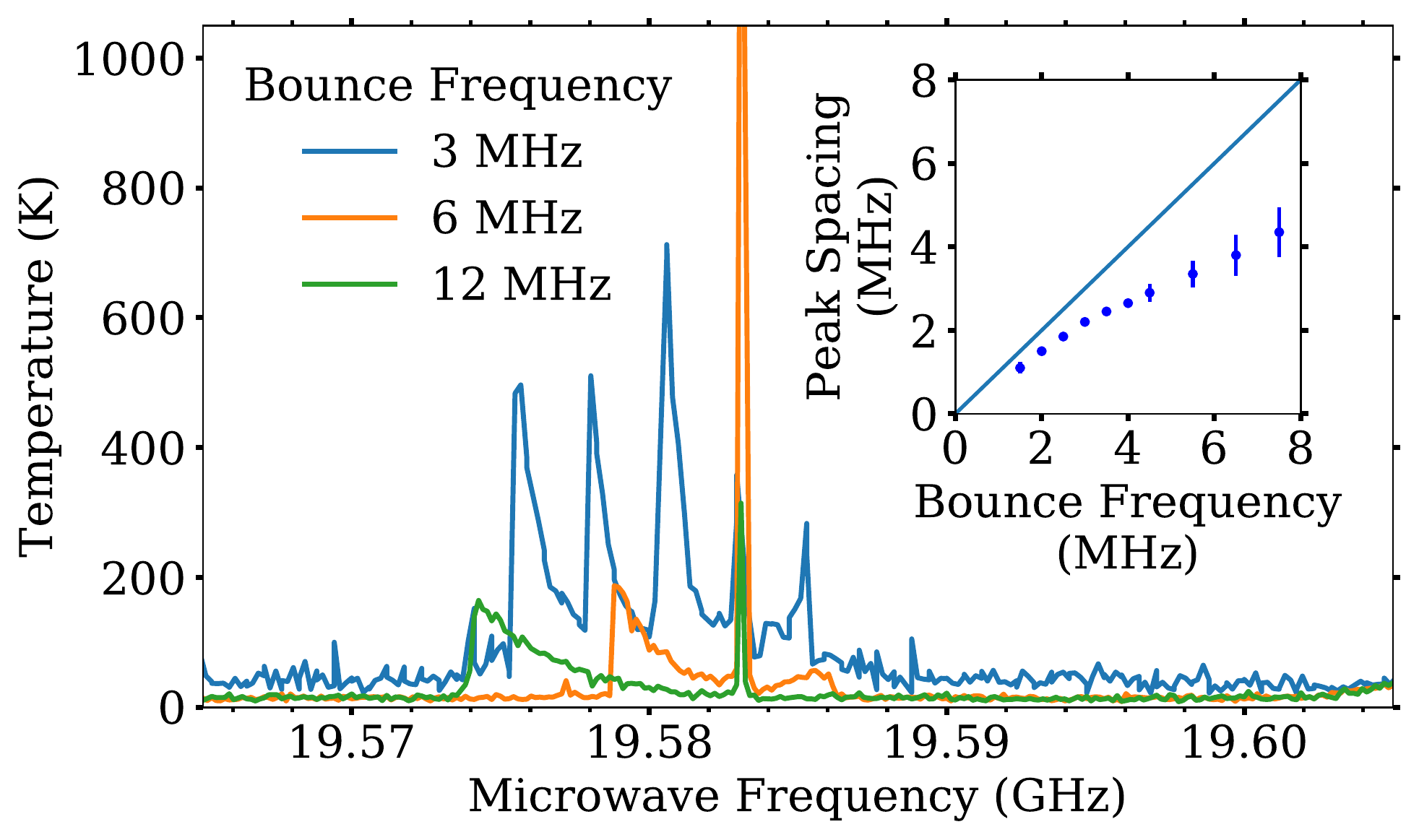}
\caption{Microwave heating scans taken with varying bounce frequencies $\omegaz/2\pi$.  (The tall orange peak rises to an off-scale value of $1750\,\mathrm{K}$).  Inset: The heating peak spacings as a function of the bounce frequency.  The line plots perfect equality between the two.}
\label{wings}
\end{figure}

If we have only a rough initial estimate of the magnetic field, as is often the case, the initial magnetometry scans must span a wide range of frequencies. Representative rough initial scans with an $\NT \approx 3\times 10^4$ plasma are shown in Fig.~\ref{wings}.

By varying the confining potential, and, hence, the bounce frequency, we can show that the peaks in Fig.~\ref{wings} come from bounce frequency sidebands.   We temporarily assume that the shape of the potential well is not significantly affected by the presence of the plasma. We use the commercial program COMSOL\cite{coms} to solve for the on-axis vacuum potential, and then approximate this numeric result with a Taylor series around the well center: $V(z)\approx V_0-k_2z^2/2-k_3z^3/6-k_4z^4/24+\cdots$.  (The well is generally near-symmetric, and $k_3\approx 0$.) Ignoring all the higher order terms, substituting the $z$ motion $z(t)=z_0\sin(\omegaz t)$, and introducing the microwave spatial dependence $\mathrm{exp}(-ikz)$, where the wavenumber $k=2\pi/\lambda$, we find that the time dependence of the electric field seen by an electron is:
\begin{multline}
\label{eq:anger}
E_-(z,t)=E_-\,\left[e^{i\omega t}e^{-ikz_0\sin(\omegaz t)} \,\right] \\
=E_-\,\left[\sum_{m=-\infty}^\infty J_m(kz_0)e^{i(\omega -m\omegaz)t}\,\right],
\end{multline}
where we have used the Jacobi-Anger identity.  Thus, the oscillating particle sees a sum of waves with frequencies $\omega-m\omegaz$ for all integers $m$. When the microwave frequency satisfies $\omega=\omegac'+m\omegaz$ for some $m$, Eq.~(\ref{eq:power}) predicts that the particle will be heated by an amount proportional to $|E_- J_m(kz_0)|^2$.

Figure~\ref{wings} shows the results of varying the electrode potentials to change $k_2$, and, hence, the bounce frequency $\fz$.  The peak spacing increases in proportion to $\fz$, particularly at lower bounce frequencies.

It is not obvious from a scan at a single bounce frequency which of the several visible peaks corresponds to the $m=0$ cyclotron frequency.  We cannot simply use the largest peak; the actual cyclotron peak in the Fig.~\ref{wings} $3\,\mathrm{MHz}$ scan is the fourth largest peak.  However, we can identify the cyclotron peak by searching for the peak that does not move as the bounce frequency is changed.

The $m\neq 0$ peaks in Fig.~\ref{wings} are much broader than the $m=0$ central peak. They are also highly asymmetric, with tails extending towards the central peak.  In the context of Eq.~(\ref{eq:anger}), this suggests that there is a distribution of bounce frequencies rather than the single bounce frequency as so far assumed.  Such a distribution would come about from the plasma self-potential flattening the vacuum well potential. Effectively, $k_2$ would become smaller and $k_4$ would become $\mathcal{O}(1/\Lp^4)$ instead of $\mathcal{O}(1/\rw^4)$, making it larger and more important.  In these circumstances, a low energy particle will have a bounce frequency tending towards zero, while a high energy particle will have a bounce frequency tending towards the original harmonic bounce frequency.  Not only will this spread the peaks in Fig.~\ref{wings}, but it will give them the appropriate asymmetric shape.  (Note that the lengths of the plasmas used in Fig.~\ref{wings} are $1\mbox{--}5$ Debye lengths long, so the well flattening is incomplete.)

For our short plasmas, $kz_0 = 2\pi z_0/\lambda \ll1$.  Thus, as the resonant frequency $\omega=\omegac'+m\omegaz$  increasingly deviates from $\omegac'$ with $|m|$, Eq.~(\ref{eq:anger}) predicts that the field strength will generally diminish as $J_m(kz_0)\propto (kz_0)^{|m|}$.  Since the heating is proportional to the square of the field strength [Eq.~(\ref{eq:power})], we would expect that the heating will likewise diminish with $|m|$.  This trend is complicated, however, by the aforementioned mode structures issues which may vary the effective incident power at different $\omega$.  Indeed, the pronounced left-right peak-magnitude asymmetry in Fig.~\ref{wings} is likely a cavity or waveguide effect. By increasing the magnetic field strength to move all peaks to the right, the peaks were partially suppressed in the band $19.585\mbox{--}19.595\,\mathrm{GHz}$, suggesting that microwaves in this band do not readily propagate to the plasma.

\section{Plasma Rotation Frequency Sidebands}
\label{sec:Rot}
\begin{figure}
\centering
\includegraphics[scale=.4]{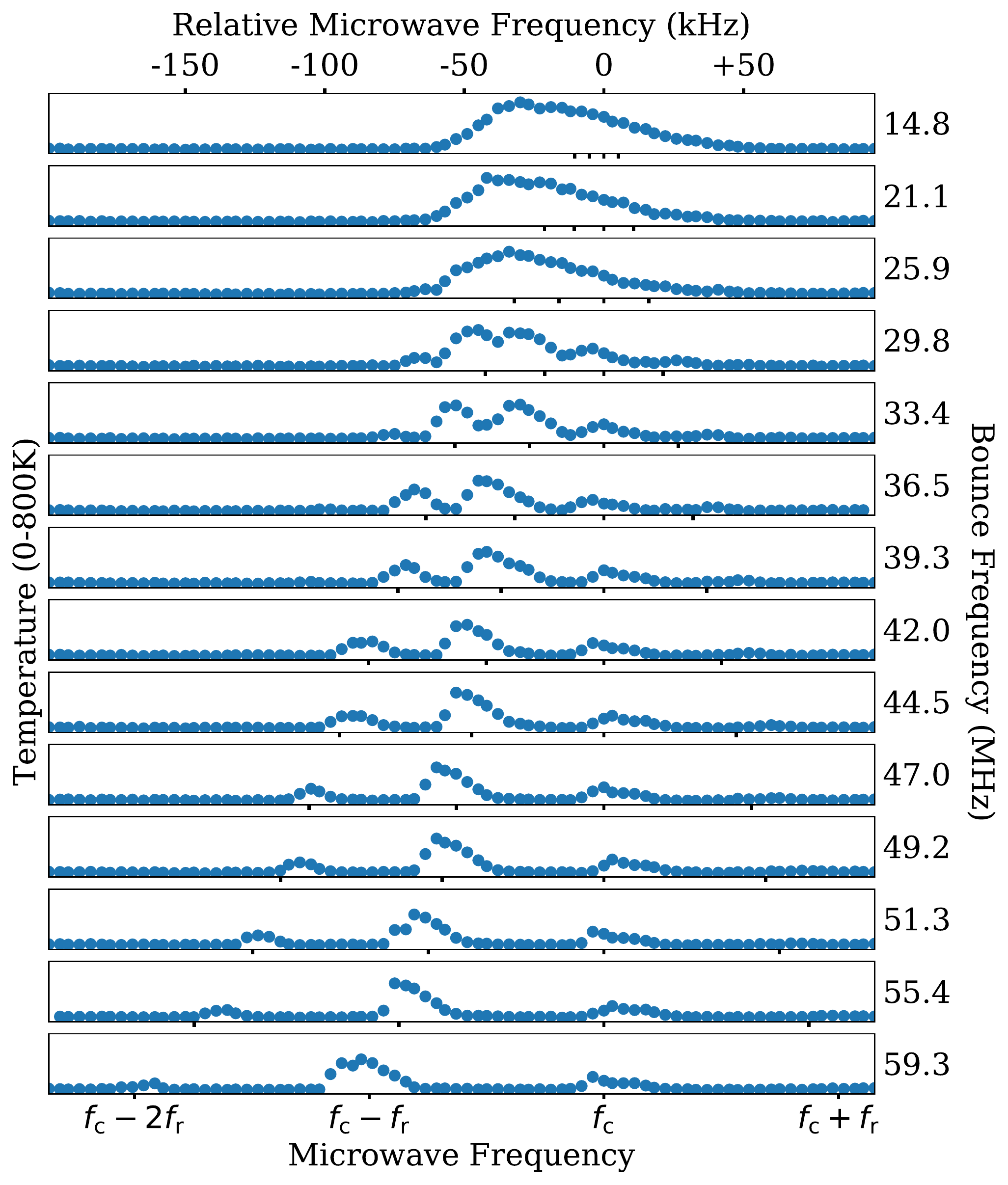}
\caption{Temperature versus microwave frequency for fourteen different bounce frequencies showing the emergence of the rotational subpeaks. The shifting frequency tick marks show the estimated cyclotron frequency with offsets of multiples of the rotation frequency. The estimated cyclotron frequency is placed in the approximate center of the non-moving subpeaks; the rotation frequency is calculated from the bounce frequency.  The baseline temperature was approximately $60\,\mathrm{K}$, and the subpeaks ranged up to about $700\,\mathrm{K}$.}
\label{RotScan}
\end{figure}

\begin{figure}
\centering
\includegraphics[scale=.4]{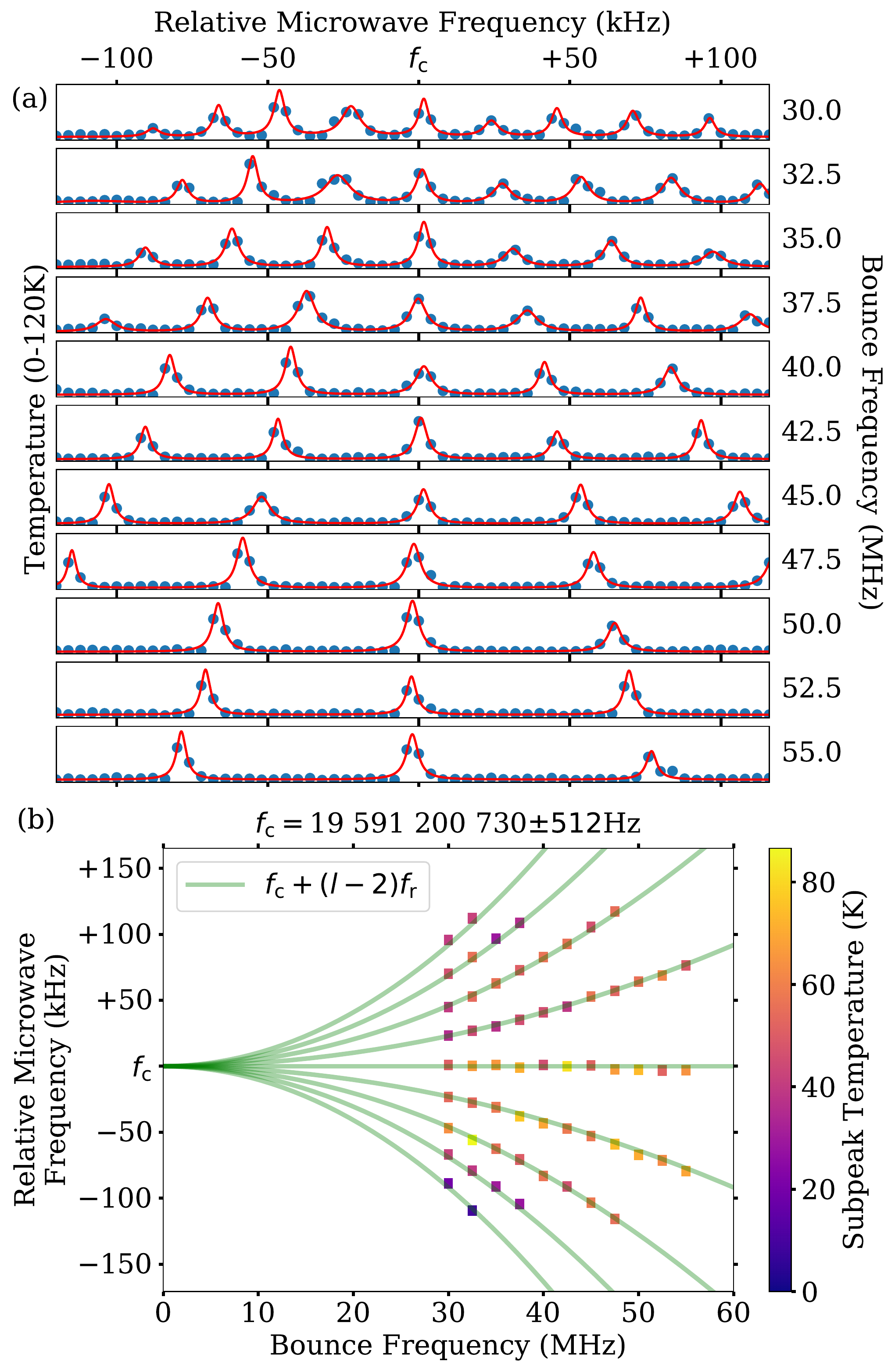}
\caption{(a) Temperature (points) and summed Lorentzian fit functions (red lines) versus microwave frequency for eleven different bounce frequencies (see Appendix~\ref{App:Fit}). The baseline temperature was approximately $15\,\mathrm{K}$, and the subpeaks ranged up to $85\,\mathrm{K}$. (b) The $61$ subpeak centers found in the Lorentzian fits then simultaneously fit to the equation $\fc+(l-2)\fr$ with the single free parameter $\fc$; for each subpeak, the $\fr$ comes from the associated bounce frequency and the $l$ is determined heuristically. The color scale indicates the measured height of the subpeaks in (a).}
\label{fig:CleanRotScan}
\end{figure}

If we narrow the microwave window to include only the central peak, and further increase the trap depth, we find that the central peak is split by a series of subpeaks separated by the rotation frequency $\omegar=\omegaz^2/2\omegac$. Figure~\ref{RotScan} shows how the subpeaks emerge as the bounce frequency is increased for a target plasma with $\NT \approx 10^4$ charges and a $\LT\approx 1\,\mathrm{mm}$ length, while Fig.~\ref{fig:CleanRotScan} shows the subpeaks in a plasma tailored to cleanly distinguish them by reducing the number of charges to $\NT \approx 1.5\times 10^3$ and a length ranging from $\LT\approx 0.1\,\mathrm{mm}$ ($\fz=55\,\mathrm{MHz}$) to $0.18\,\mathrm{mm}$ ($\fz=30\,\mathrm{MHz}$).

Rotation frequency dependent cyclotron resonance phenomena have been explored theoretically,\cite{davi:90,goul:95,dubi:13} and confirmed experimentally, in electron\cite{goul:92} and multi-species ion\cite{sari:95,affo:15} nonneutral plasma systems.  However, it is not obvious that these results are completely applicable to our experiment.  The prior work modeled/employed long plasmas with many particles where the trapping fields have only small effects on the rotation, and quasi-electrostatic drives.  We employ short plasmas with few particles where the trapping fields dominate the rotation, and a fully electromagnetic drive.

As we do not yet have a fully appropriate model of our system, we will present a single-particle model of the resonant structures.  The model begins by generalizing the definition [Eq.~(\ref{eq:Eoriginal})] of the applied microwave electric field to include rotational modes.  Thus, in cylindrical coordinates $({\hat r},{\hat \theta})$,
\begin{equation}
\label{eq:Ecyl}
\begin{split}
\mathbf{E}=&\left[E_+(r,\theta)({\hat r}+i{\hat\theta}) + E_-(r,\theta)({\hat r}-i{\hat\theta})\right]\\
&\times\exp[i(-l\theta+\omega t-kz)],
\end{split}
\end{equation}
where $l$ is the rotational mode number.  For notational simplicity, we will generally suppress the dependencies on $(r,\theta)$ or $(x,y)$ of the $E$ fields, and the $-kz$ dependence.  Changing the unit basis back to $({\hat x},{\hat y})$ yields
\begin{equation}
\label{eq:Exy}
\begin{split}
\mathbf{E}=&E_+({\hat x}+i{\hat y})\exp[i(-[l+1]\theta+\omega t)]\\
              &+ E_-({\hat x}-i{\hat y})\exp[i(-[l-1]\theta+\omega t)].
\end{split}
\end{equation}
As before, only the $E_-$ term will be resonant.

We next assume that the electron motion consists of a gross ($r \lesssim\rp$), slow rotation around the trap axis, such that the angular position of the electron is well approximated by $\theta=\omegar t$, combined with a small ($\rL/\rp \lesssim 0.001$, where $\rL$ is the Larmor radius), fast rotation at the cyclotron frequency $\omegac'$ [see Eqs.~(\ref{Eq:cyclotron}) and (\ref{Eq:rotation})]. Then the resonant field can be written as
\begin{equation}\label{eq:Eres}
\mathbf{E}=E_-({\hat x}-i{\hat y})\exp[i(-[l-1]\omegar+\omega) t].
\end{equation}
Since $\omegac'\gg\omegar$, the electron velocity follows $\exp[i\omegac' t] = \exp[i(\omegac-\omegar) t]$, and the resonance condition gives a shift from the cyclotron frequency $\omegac$ of
\begin{equation}
\label{eq:res}
\delta\omegac = \omega-\omega_c = (l-2){\omegar}.
\end{equation}
This shift is closely related to the shifts found in prior work,\cite{davi:90,goul:95,goul:92,sari:95,dubi:13,affo:15} with the plasma self-rotation substituted for our magnetron rotation.  One significant difference between our work and prior work is that we do not find that $l$ is restricted to non-negative integers.  Experimentally, $l=-2$ and $-1$ subpeaks are observable in Fig.~\ref{fig:CleanRotScan}. Such counter-rotating modes were not predicted by prior theories, nor were they observed in prior experiments.  In Appendix~\ref{App:Negl} we show that this is a consequence of the choice of drive; negative $l$ modes are not allowed for quasi-electrostatic drives, but they are allowed, albeit at lesser magnitude, for our electromagnetic drive.

\section{Magnetometry}
\label{sec:Mag}
From Eq.~(\ref{eq:res}), we see that the $l=2$ subpeak of the $m=0$ main peak is independent of the bounce and rotation frequencies, and, hence, is a good candidate for the ``true'' cyclotron peak. Often, we have already identified the $l=2$ subpeak from the ensemble of peaks by a prior bounce frequency study or because it has been tracked through time or small variations in the plasma location.  In this case, we can measure the magnetic field by finding the central frequency of the $l=2$ subpeak with a single microwave frequency scan at any bounce frequency and density where the subpeak is clearly identifiable.  Many such scans are shown in Fig.~\ref{fig:CleanRotScan}.

In Appendix~\ref{App:Errors}, we estimate the errors in our measurement. Known shifts of the $l=2$ subpeak from the true cyclotron frequency come from environmental effects ($\sim +0.3\,\mathrm{kHz}$), plasma charge effects ($\sim +1\,\mathrm{kHz}$), and temperature effects ($\sim -0.2\,\mathrm{kHz}$), for a net shift of $\sim +1.1\,\mathrm{kHz}$.  However, these shifts are not yet well enough understood to warrant simply subtracting them from our observed answer, and we choose to keep them as systematic errors.  In addition, there is a statistical uncertainty in locating the resonance subpeak of $\pm 2.0\,\mathrm{kHz}$. Adding the net shift and the statistical uncertainty gives error bounds of $-3.1\,\mathrm{kHz}$ to $+0.9\,\mathrm{kHz}$ on the cyclotron frequency.  Taking the larger bound as our uncertainty, we get a systematic error of less than  $\pm 1\,\mathrm{ppm}$ ($\pm 3.1\,\mathrm{kHz}$ or $\pm 0.17\,\mathrm{ppm}$).

We can increase the precision of our field estimate by simultaneously analyzing all identifiable $l$ subpeaks, and mapping them back to zero bounce frequency following the procedure described in Appendix~\ref{App:Fit}. For the data in Fig.~\ref{fig:CleanRotScan}, this yields a precision of $26\,\mathrm{ppb}$, essentially eliminating the statistical error. However, because the environmental drifts are larger over the time required to collect all the data in Fig.~\ref{fig:CleanRotScan}, as opposed to just one of the bounce frequency scans, the absolute error does not significantly improve.

\subsection{Spatial Field Maps}
\begin{figure}
\centering
\includegraphics[scale=.4]{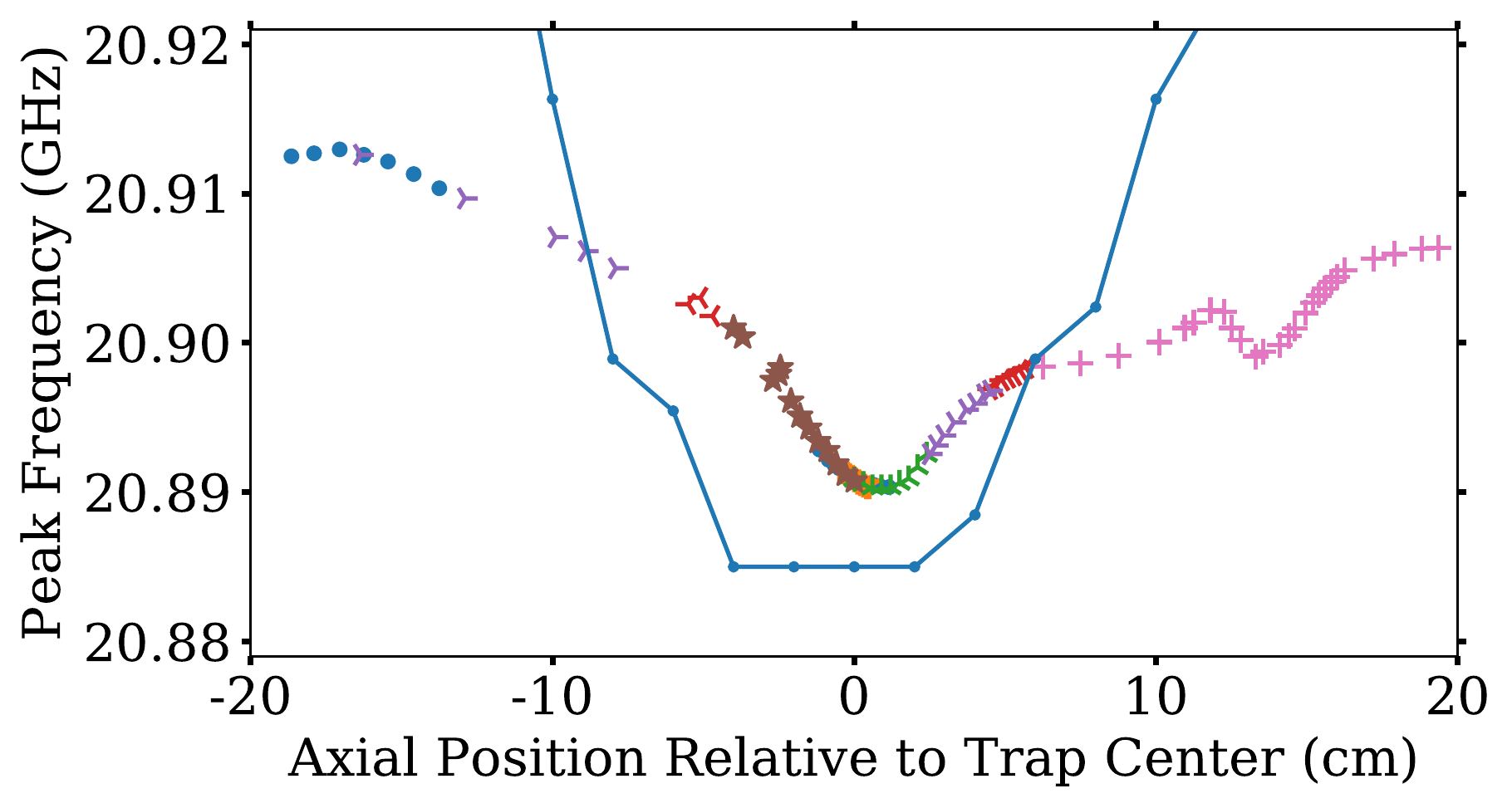}
\caption{Plot of the cyclotron frequency versus axial position. The different datapoint colors and shapes refer to different $k_2$ well coefficients. The resulting bounce frequency range is $\fz=2\mbox{--}10\,\mathrm{MHz}$. The blue curve is a scaled version of the measurement recorded in the manual for our Oxford magnet (taken at $6\,\mathrm{T}$). No attempt was made to compensate for the possible axial offset of the Penning-Malmberg trap.}
\label{map}
\end{figure}

One application of ECR magnetometry is to map the magnetic field $|B|$ along the axis of a trap. A plot of our trap's solenoid field is shown in Fig.~\ref{map}. The figure compares our measurements to those taken by the manufacturer (in the absence of the trap vacuum structure) many years ago.  To make this detailed map, we need to move the target plasma along the trap axis in steps that are less than the trap electrode lengths.  We do this by applying asymmetric voltages to the electrodes forming the electrostatic well barriers.  This technique can move the well center continuously at the expense of limiting the range of well depths ($k_2$) that we can apply.  However, since the field gradients are small, we do not need to thoroughly scan $k_2$ as the cyclotron peak  can be tracked as the plasma is moved.

\subsection{Measurements Near a Magnetic Saddle}
One scheme to measure the effect of gravity on antihydrogen relies on diamagnetic forces from the gradients generated by magnetic mirror coils.\cite{amol:13,zhmo:13}  In this scheme, it is critical to control and measure the field at the saddle points in the axial center of the mirrors. The field must be known to better than $1\,\mathrm{G}$ to measure the sign of gravity, and about ten times better to measure the gravitational acceleration to 1\%.

ECR magnetometry can be used to measure the field, but the technique is complicated by the field inhomogeneities near the saddle.  For a $1\,\mathrm{T}$ mirror, a $1\,\mathrm{G}$ measurement is at the $100\,\mathrm{ppm}$ level.  A $4.5\,\mathrm{cm}$ radius mirror would then demand that the target plasma be less than about $1\,\mathrm{mm}$ in length and radius, and contain one to ten thousand electrons. Such plasmas can be made by our reservoir technique.  The inhomogeneities will smear the heating peaks, and would make it difficult to distinguish the rotational resonances.  It would be possible and necessary, however, to distinguish the bounce resonances.

The ECR target plasma would have to be axially positioned to cover the saddle.  As with the field map, this can be accomplished by driving the trap electrodes with asymmetric potentials.  Because the trap electrodes and the corresponding mirror coil cannot be perfectly registered due to construction issues, one would have to axially search for the saddle center. The center can be identified as the point which yields the highest magnetic field.

\subsection{Measurements in a Gradient}
Measurements made in a magnetic gradient would be limited by many of the same concerns found for saddle measurements. The accuracy and precision of such measurements would be limited by the plasma sample size. At $10\,\mathrm{K}$ and with a bounce frequency of $50\,\mathrm{MHz}$, the minimum plasma length would be about $0.1\mathrm{mm}$.  Thus, for a $1\,\mathrm{G}$ measurement, the field gradient cannot be stronger than $10\,\mathrm{G}/\mathrm{mm}$. Registration issues would also be important, and here one would not have the benefit of the effective fiducial found at a saddle center.

\subsection{Low and High Field Measurements}
The measurements reported here were taken in the vicinity of $0.7\,\mathrm{T}$.  In other experiments, we have measured fields of $0.17\,\mathrm{T}$ with roughly the same $\sim 1\,\mathrm{ppm}$ accuracy.  Measurements at such lower fields are more difficult because of the lack of cyclotron cooling. The initial plasma temperatures before microwave illumination were substantially hotter in this lower field: about $500\,\mathrm{K}$ vs.\ about $15\,\mathrm{K}$ in Fig.~\ref{fig:CleanRotScan}.  Lower temperatures could be obtained by using cavity resonances,\cite{hunt:18} but this might confuse the mode identification.

Our magnetometry technique relies on rapid thermalization of the target plasmas' parallel and perpendicular temperatures.  At low temperatures and high fields, the plasma enters the strongly magnetized regime in which thermalization is inhibited by O'Neil's adiabatic invariant.\cite{onei:83}  At sufficiently high fields, this might require that the technique be adjusted to keep the baseline temperatures above a reasonable thermalization time threshold.  Otherwise, assuming appropriate microwave sources are available, measurements at high fields present no new problems.

\section{Conclusion}
\label{sec:Conc}
We have described an improved technique for measuring the magnetic field magnitude in a Penning-Malmberg trap employing ECR heating.  The technique is based on a new method for rapidly generating very small plasmas, and on the unambiguous identification of the unshifted cyclotron peak in the presence of a rich resonance structure. Measurements with absolute accuracies better than $1\,\mathrm{ppm}$ can be obtained in less than one minute, an improvement of over a factor of $\sim5 0$ in accuracy and $\sim 10$ in time from our previous practice.\cite{amol:14b} Repeated measurements with varying confinement well parameters can result in precisions at the $26\,\mathrm{ppb}$ level, although the absolute accuracy does not significantly improve due to increased environmental drifts.  This constitutes an improvement by a factor of more than one thousand over previous practice.

Prior experiments in Penning-Malmberg traps at other facilities were at the $1\%$ level,\cite{goul:92} the $100\,\mathrm{ppm}$ level,\cite{sari:95} and the $200\,\mathrm{ppm}$ level.\cite{affo:15}  Fourier transform ion cyclotron resonance (FTICR) mass spectroscopy devices typically determine masses to the few $\mathrm{ppm}$ level, from which the magnetic field can be backed out with similar precision.  Much improved precision can be obtained in these devices by comparing masses,\cite{savo:11} but this does not yield the magnetic field.  Highly specialized Penning traps, often working with a single particle, can measure the magnetic field at the few $\mathrm{ppb}$ level.\cite{smor:18}

Plasma-based precision ECR magnetometry in Penning-Malmberg traps is a new field of study.  Its limitations and ultimate precision need to be further explored with experiments, theory, and simulations.  For example, fully understanding the plasma charge shifts would allow us adjust the observed $l=2$ frequency and remove the dominant systematic error.  A better understanding of the lineshape, and more closely spaced measurements, would allow us to reduce the statistical errors.

This work was motivated by experiments exploring fundamental physics with antihydrogen.  Magnetic field errors are the dominant error source for the ALPHA collaboration's planned gravity experiments,\cite{amol:13,zhmo:13,hami:14} and accurate field measurements undergird ALPHA's understanding of the systematic errors in ongoing 1S--2S,\cite{ahma:18a} 1S--2P,\cite{ahma:18b} hyperfine,\cite{ahma:17b} Lamb shift, and laser cooling measurements.

\begin{acknowledgements}
We thank A.~Charman and N.~Evetts for their comments on this work. EDH acknowledges encouraging and helpful email exchanges with C.~Carruth while implementing the SDREVC protocol at Berkeley. This work was supported by the DOE OFES and NSF-DOE Program in Basic Plasma Science.
\end{acknowledgements}
\appendix

\section{Plasma Expansion from EVC and Extraction}
\label{App:ResExpansion}
Target plasma extraction and EVC cause the plasma radius to increase because of angular momentum conservation.\cite{onei:80} In Ref.~\onlinecite{andr:10}, it was shown that EVC would increase the average radius proportional to $\sqrt{\NI/\NF}$, where $\NI$ and $\NF$ are the initial and final plasma particle numbers.  This result relies on the assumption that all the escaped particles leave on the $r=0$ axis, and that the plasma never leaves, or, subsequent to EVC, reenters global thermal equilibrium.

For extraction of particles from the reservoir (see Sec.~\ref{sec:extr}), the derivation is more complicated as particles leave the reservoir at radii up to the radii of the target plasmas.  Recall that the total angular momentum of a strongly magnetized nonneutral plasma is given by\cite{onei:80}
\begin{equation}
P_\theta\propto \int |r|^2\rho(\mathbf{r})d^3\mathbf{r}\approx \frac{1}{2} \NTot\rp^2,
\end{equation}
where $\rho(\mathbf{r})$ is the charge density.  The second equality comes from approximating the plasma as having $\NTot$ charges uniformly distributed out to plasma radius $\rp$. If the reservoir begins with $\NRi$ electrons out to radius $\rRi$, and we draw $\NT$ electrons into a target plasma of radius $\rT$, we find that the reservoir's final radius after each extraction is:
\begin{equation}
\label{resev}
\rRf=(1+\alpha)\rRi\sqrt{\frac{\NRi-\NT(\rT/\rRi)^2}{\NRi-\NT}}>\rRi,
\end{equation}
where $\alpha$ is a fit anomalous expansion factor discussed later.  Note that $\NT$ will decrease and $\rT$ will increase after the final evaporative cooling step~f.

Assuming that the extraction steps b--d are done sufficiently slowly that the plasma is always in thermal equilibrium as the target and reservoir plasmas are separated, the plasma will be shear-free\cite{pras:79} until the moment of separation.  If we further assume that the local magnetic field is invariant, then the interior plasma density and the $r=0$ potential must be constant;\cite{faja:03}  the plasma will satisfy these conditions by adjusting its radius.\cite{peur:90}  As we impose the condition that the vacuum potential in the target region is less positive than in the reservoir region (see Sec.~\ref{sec:extr}, step~b), the target radius will thus be less than the reservoir radius.  This smaller radius is visible in Fig.~\ref{fig:reservoir_extraction}b,  and establishes the inequality in Eq.~(\ref{resev}).

\begin{figure}
\centering
\includegraphics[scale=.4]{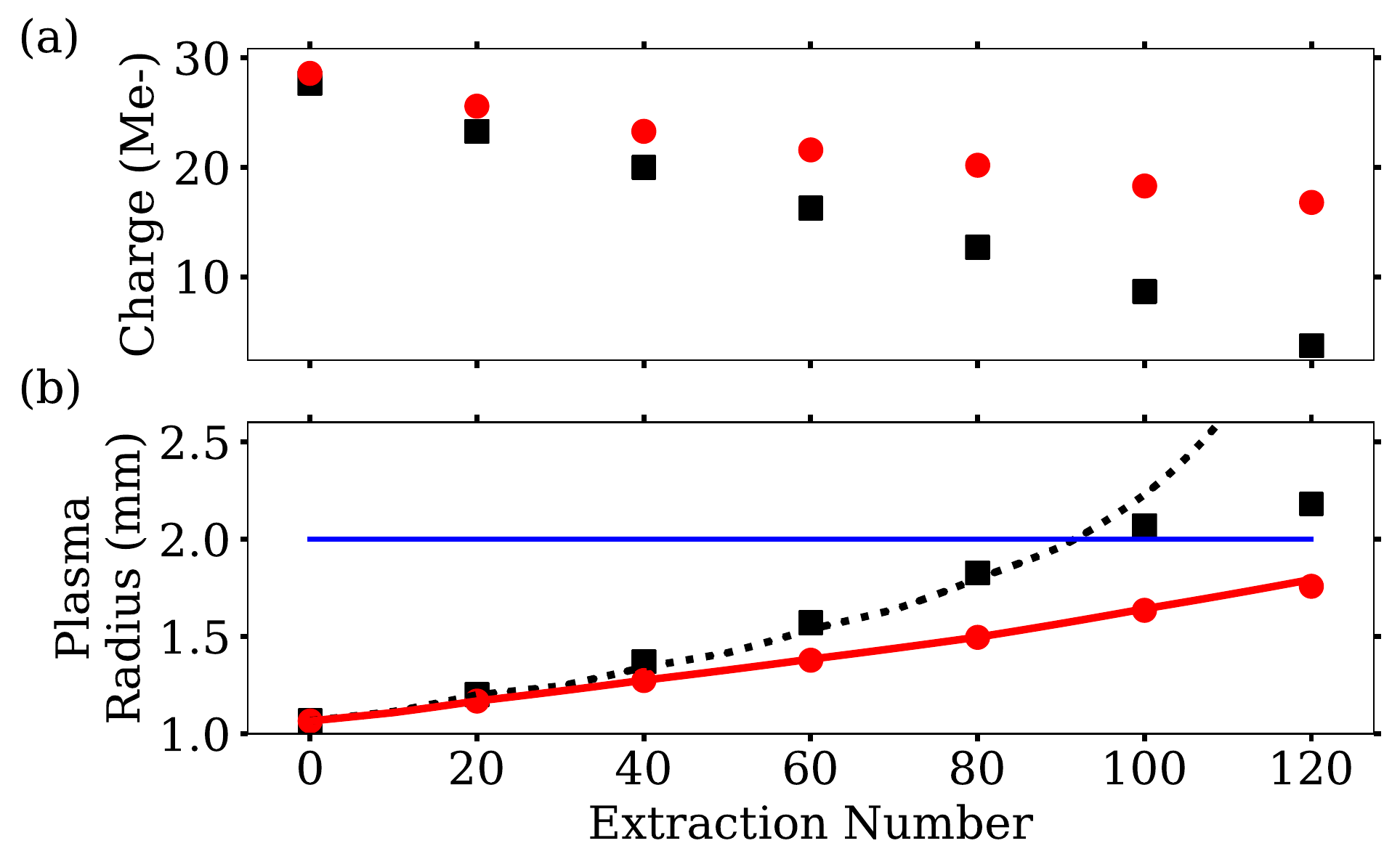}
\caption{
Reservoir charge (a) and radius (b) as a function of the extraction number while using bias voltage steps of 25mV (red circles) and 50mV (black squares). The red solid and black dotted lines show the predicted radii from Eq.~(\ref{resev}), iteratively applied, using the target plasma data in Fig.~\ref{Fig:TargetPlasmas}, with an expansion parameter $\alpha$ fitted to the data. The blue horizontal line in (b) shows the plasma radius which would completely fill the phosphor screen if it were perfectly centered. The black squares above or near this line are, therefore, not reliable and are not used in fitting $\alpha$. The rate $\alpha =1.00252/\text{extraction}$ fits both step sizes, and is a $35\%$ correction to Eq.~(\ref{resev}) after $120$ extractions.}
\label{Fig:ResEvolution}
\end{figure}

In Fig.~\ref{Fig:ResEvolution}, we show how the reservoir evolves over time. The measured reservoir radius increases more quickly than predicted by Eq.~(\ref{resev}). This suggests the presence of additional sources of plasma expansion. Note that the potentials applied during the extraction of a target plasma resemble the ``squeeze'' potentials employed by the UCSD group to study transport effects.\cite{dubi:11,kaba:14,dubi:17} They explain that squeeze-driven transport comes from particles quasi-trapped on one or the other side of the squeeze.  These particles drift for many orbits before recrossing the squeeze separatrix. If we heuristically modify Eq.~(\ref{resev}) to incorporate a constant proportional expansion rate per extraction $\alpha$ from this effect, we obtain the fit lines plotted in Fig.~\ref{Fig:ResEvolution}.

\section{Negative $l$ Modes}
\label{App:Negl}
References~\onlinecite{davi:90,goul:95} derive an equation analogous to Eq.~(\ref{eq:res}) under the assumption that the driving electric field has no dependence on $z$: i.e.\ that the axial wavenumber $k$ is zero.  Such drives can be produced by rotating voltages applied to azimuthal sectors on the trap wall, yielding the quasi-electrostatic potentials
\begin{equation}
\label{eq:ElectrostaticPot}
\Phi(r,\theta)=\Phi_0 r^{|l|}\exp[i(\omega t-l\theta)].
\end{equation}
Taking the negative gradient of Eq.~(\ref{eq:ElectrostaticPot}), and using the relations $E_+=(1/2)(E_r-iE_\theta)$ and $E_-=(1/2)(E_r+iE_\theta)$, yields the electric fields
\begin{equation}
\label{eq:ElectrostaticE}
E_\pm = -\frac{|l|}{2}\left(1\mp\sgn(l)\right)\Phi_0 r^{|l|-1}\exp[i(\omega t-l\theta)].
\end{equation}
For $l>0$, $|E_-|>0$ and $|E_+|=0$.  Since $|E_-|$ is resonant, this drive will interact with the electrons and cause a heating subpeak.  However, if $l<0$, $|E_+|>0$ and $|E_-|=0$. Since $|E_+|$ is not resonant, the drive will not cause a heating subpeak.

Experimentally, Refs.~\onlinecite{goul:92,affo:15,sari:95} use azimuthal sectors of their trap wall to drive their plasmas.  An example of such a sector is shown on the wall between the Target and Reservoir plasmas in Fig.~\ref{stack}.  Finite length sectors produce fields with some $z$ dependence, and cannot be perfectly represented by Eq.~(\ref{eq:ElectrostaticE}).  Nonetheless, Eq.~(\ref{eq:ElectrostaticE}) captures the basic transverse properties of the field near the axial center of the sector.  Hence, negative $l$ modes are not expected theoretically, or observed experimentally, for the configurations explored in Refs.~\onlinecite{davi:90,goul:95,goul:92,affo:15,sari:95}.

For completeness, note that the $l=0$ mode can be excited by quasi-electrostatic drives applied at the end of a finite length plasma, and was observed by Affolter et al.\cite{affo:15}

We drive our plasmas with an electromagnetic wave. Using standard electromagnetic theory for a TE$_{lp}$ wave, for example, shows that
\begin{equation}
\label{eq:EM wave}
\begin{split}
E_r(r,\theta,z;l,p) =& C_{lp}^\pm l \frac{J_l(\rho_{lp}')}{\rho_{lp}'}\exp[i(\omega t-l\theta-kz)]\\
E_\theta(r,\theta,z;l,p) =& -iC_{lp}^\pm J_l'(\rho_{lp}')\exp[i(\omega t-l\theta-kz)]\\
\end{split}
\end{equation}
where $\rho_{lp}'=j_{lp}' r/\rw$, $j_{lp}'$ is the $p$th root of $J_l'$, and $C_{lp}^\pm$ is a mode-dependent constant.  Then
\begin{equation}
\label{eq:EM circ}
\begin{split}
E_\pm(r,\theta,z;l,p) =& \mp\frac{C_{lp}^\pm}{2}\left[J_l'(\rho_{lp}') \mp l\frac{J_l(\rho_{lp}')}{\rho_{lp}'}\right]\\
                        &\times\exp[i(\omega t-l\theta-kz)].
\end{split}
\end{equation}
The resonant $E_-$ terms no longer vanish for negative $l$.  In agreement with our observations, heating subpeaks may exist for both positive and negative $l$.  As we do not control the modes present in our trap, the power directed towards positive and negative $l$ modes may be different.  This effect is partially masked, however, by the fact that our heating subpeaks are often ``saturated'' in the sense that increasing the microwave power does not increase the final temperature of the subpeaks.

\section{Magnetometry Errors}
\label{App:Errors}
\subsection{Signal Generator Errors}
\label{App:ErrorsSigGen}
Our determination of the magnetic field is no more accurate than the calibration of our HP~8673d signal generator. On the time scale of our measurements, this signal generator has frequency drifts at the $\mathrm{ppb}$ level, which we may neglect.  However the absolute accuracy of the frequency is an unspecified parameter, and our signal generator has not been calibrated recently.  This introduces an unknown, and possibly large, error into our measurements.  This error is not intrinsic to our measurement technique, and is easily remedied by using a calibrated frequency source.  Thus, we choose not to report it in our error estimates.

\subsection{Peak Location Errors}
The location of any individual peak, including the $l=2$ subpeak, in a single bounce frequency scan, can be identified to within the microwave frequency separation of $4\,\mathrm{kHz}$ [i.e.\ $\pm 2\,\mathrm{kHz}$ ($\pm 0.1\,\mathrm{ppm}$)]. At the expense of increased scan time, this uncertainty would probably decrease with a tighter scan.

\subsection{Environmental Errors}
With our solenoidal magnet in persistent current mode and with our electron source on, we observe an upward frequency drift of all the subpeaks.  While small, this drift is too large to be the unavoidable decay of the persistent current.  Nor is it a residual field effect (the rearrangements of currents in the magnet wire's superconducting filaments\cite{voll:03}). These effects should have died out as the magnet has typically been in persistence mode for several days.  Furthermore, these drifts are reversible when the electron source is turned off.

The drifts appears to be caused by the heat generated by the electron source; we observe a roughly linear relationship between the drift and the length of time the source has run continuously.  Though largely independently cooled, the source, which dissipates approximately $1\,\mathrm{W}$, is located inside the solenoid and some of the heat that it generates couples to the solenoid bore and to the electrodes through both radiation and conduction.

With the electron source running continuously, this drift is on the order of $10\,\mathrm{kHz}$ in ten minutes and is reversible on a somewhat longer timescale when the source is turned off. We have measured the same drift (to one decimal place) at three locations in the electrode stack separated by a total of $\sim 23\,\mathrm{cm}$.  We have also repeated these measurement at  $0.16\,\mathrm{T}$; this field is four times lower than used in the rest of this paper. We then observed a factor-of-four reduction in the drift with comparable electron source run-time. These measurements suggest that the drift comes from a heat-induced change in the solenoidal field.

We also obtain comparable reversible frequency drifts by temporarily pressurizing the liquid helium (LHe) reservoir used to cool the solenoid by approximately $0.1\,\mathrm{bar}$. This would raise the LHe temperature by approximately $0.1\,\mathrm{K}$.  It would also readjust the stresses in the mechanical supports of the solenoid, and possibly shift its position.  We observe similar reversible effects when we fill the solenoid's LHe reservoir.

The trap vacuum is completely decoupled from the LHe reservoir. Thus, if these source and LHe pressure drifts originate from the same cause, a pressure increase in the trap appears to be ruled out.  There could be many affects at this level that are caused by thermal expansion. Conceivably, there could be effects from temperature-induced changes in the magnetic properties of the materials in the solenoid and trap.

Regardless of the mechanism, we believe that we are measuring real drifts in the magnetic field.  These drifts would likely not occur in a device in which the electron source was well removed from the solenoid.  Nonetheless, we conservatively classify these drifts as a systematic error and strive to minimize them by running the electron source only long enough to capture the reservoir plasma.  For a scan at a single bounce frequency (one curve in Fig.~\ref{fig:CleanRotScan}) the drift is about $0.3\,\mathrm{kHz}$ ($0.015\,\mathrm{ppm}$); to take all the data in the figure, the drift is about $3\,\mathrm{kHz}$ ($0.15\,\mathrm{ppm}$).  To minimize the effect of this drift, we interleaved acquiring the data following the pattern: 1 (lowest bounce frequency), 3, 5, 7, 9, 11 (highest bounce frequency), 2, 4, 6, 8, 10.

\subsection{Plasma Effects}
The frequency of the stationary $l=2$ subpeak is a strong candidate for a measurement of the exact cyclotron frequency. However, we have so far assumed that the resonant peak structure comes solely from the incident microwave frequencies and the details of the electrostatic well confining the plasma, and is not sensitive to the parameters of the plasma itself.  We next consider these effects.

\subsubsection{Theoretical Limits on Plasma Charge Effects}
We do not have a theory for short plasmas subject to electromagnetic perturbations.  As mentioned earlier, there are important deficiencies, most notably the lack of negative $l$ modes, in applying an electrostatic perturbation theory to our experiment. Nonetheless, it is illustrative to modify the ``standard'' electrostatic theory \cite{davi:90,goul:95,sari:95,dubi:13,affo:15} to cover our case of short plasmas where the rotation is primarily driven by the wall potential.  We start with Davidson's electrostatic dispersion relation for a multispecies, infinite length plasma column,\cite{davi:90} reduced to the one species case:
\begin{equation}
\label{eq:dispersion}
0=1-\frac{\omegap^2[1-(\rp/\rw)^{2l}]}{2(\omega-l\baromegar)[(\omega-l\baromegar)+(2\baromegar-\omegac)]},
\end{equation}
where $\baromegar=\omegar+\omegas$ is the total rotation frequency: the sum the magnetron rotation frequency $\omegar$ and the plasma self-charge rotation frequency $\omegas$. Solving Eq.~\ref{eq:dispersion} for frequencies close to the cyclotron frequency, $\omega=\omegac+\delta\omega$, yields
\begin{equation}
\label{eq:dispersion_I}
\delta\omega=\left[(l-2)+\frac{\omegap^2}{2\omegac\baromegar}[1-(\rp/\rw)^{2l}]\right]\baromegar,
\end{equation}
where we have assumed that $\omegac$ is much greater than $\delta\omega$ and $l\baromegar$.

The self-rotation frequency for an infinite-length nonneutral plasma is given by $\omegas=\omegap^2/2\omegac$.  However, the derivation of this formula assumes that the plasma flattens the axial potential and there is no interior axial electric field.  Our pancake-shaped target plasmas are not cold enough to attain this regime, and most of the electric field from the plasma charge is ``wasted'' out axially. Consequently, the radial electric field, and hence the self-rotation frequency, are both reduced by some factor ${\bar G}$.  From numeric potential calculations, we find that ${\bar G}\sim 0.15$. (This factor is similar, but not identical, to the analytically calculated factor $G$ in Jeffries, et al.\cite{jeff:83}) The revised formula $\baromegas={\bar G}\omegap^2/2\omegac$ can then be immediately inserted into the expression for $\baromegar$, which, for the parameters of Fig.~\ref{fig:CleanRotScan}, increases $\baromegar$ by approximately $2\%$.

In addition to shifting the rotation frequency, the plasma charge affects the cyclotron resonance through the $\omegap^2$ term in Eq.~(\ref{eq:dispersion_I}).  This term produces\cite{goul:92,dubi:13} Bernstein-like modes. Such modes are driven by the radial electric fields generated by the modes' self-induced charge density perturbations.  As with the rotation frequency, the pancake-like plasma profile reduce the modes' radial electric fields. We will estimate these effects using the same factor ${\bar G}$ used to estimate $\baromegas$.

Using $\baromegas$ for $\omegap^2/2\omegac$ in Eq.~(\ref{eq:dispersion_I}) gives
\begin{equation}
\label{eq:dispersion_II}
\delta\omega=\left[(l-2)+\frac{\baromegas}{\baromegar}[1-(\rp/\rw)^{2l}]\right]\baromegar,
\end{equation}
which reduces to
\begin{equation}
\label{eq:dispersion_III}
\begin{split}
\delta\omega &=(l-2)\omegar + [l-1-(\rp/\rw)^{2l}]\baromegas,\\
            &\approx (l-2)\omegar + (l-1)\baromegas \quad\quad\text{if}\ l\ne 0,
\end{split}
\end{equation}
where the simplification on the last line is justified when image charges can be neglected (i.e.\ when $\rp/\rw\ll 1$).

When image charges can be so neglected,  Wineland and Dehmelt\cite{wine:75a} show that self-charge interactions neither shift nor broaden the cyclotron resonance for the quasi-spatially-uniform $l=1$ mode.  For this mode, Eq.~(\ref{eq:dispersion_III}) simplifies to $\delta\omega = -\omega_r$, which is indeed independent of the plasma charge.  This result supports our use of the same ${\bar G}$ for both the self-rotation frequency and for the Bernstein modes.  A more accurate treatment might require different Bernstein mode ${\bar G}_l$ for every $l\neq 1$.

For the $l=2$ mode, Eq.~(\ref{eq:dispersion_III}) reduces to a shift $\delta\omega=\baromegas$ when, as is the case here, $\rp/\rw$ is small.  Our numerical studies yield $\baromegas/2\pi\approx 1\,\mathrm{kHz}$, or a $0.05\,\mathrm{ppm}$ systematic error from the plasma charge shift.  This shift could be reduced by decreasing the number of electrons in the target plasma.  We know of no reason that we could not, for instance, adequately measure the temperature of a plasma with $100$ electrons, provided that the collision frequency is still adequate to redistribute the perpendicular energy gained through microwave illumination into the parallel energy measured by our temperature diagnostic.

Equation~(\ref{eq:dispersion_II}) can also be rewritten as
\begin{equation}
\label{eq:dispersion_IV}
\delta\omega=\left[(l-2)+\delta(1-(\rp/\rw)^{2l})\right]\baromegar,
\end{equation}
where $\delta=\baromegas/\baromegar$ is the ratio of the self-charge rotation frequency to the wall potential rotation frequency.
If we reinterpret $\delta$ as a species fraction, Eq.~(\ref{eq:dispersion_IV}) is identical to the electrostatic dispersion equation previously derived for long multispecies plasmas.\cite{davi:90,goul:95,sari:95,dubi:13,affo:15}

For completeness, note that there are effects on the magnetron rotation frequency $\omega_r$ originating from changes in the confining potential anharmonicities sampled by the plasma as its shape changes with its charge.  These effects are small, and since the $l=2$ shift is independent of $\omegar$, we do not believe that these effects affect our measurement.

\subsubsection{Experimental Limits on Plasma Charge Effects}
We obtained a limited dataset (not shown), which measured the $l=2$ shift at $\NT=10^3$ and $\NT=10^5$.  This dataset bounds the plasma charge dependent frequency shift for  $\NT=1.5\times 10^3$ at approximately $0.35\,\mathrm{kHz}$ ($0.02\,\mathrm{ppm}$) in our apparatus. This is smaller than the theory bound for the plasma charge shift; to be conservative, we will use the theory bound to calculate the systematic error from plasma charge effects.

We have not been able to find results on other experiments that closely match our setup, but there are some results that are perhaps relevant.   Affolter et al.,\cite{affo:15} for instance, finds a dependence on $\omegas$ similar to that given by Eq.~(\ref{eq:dispersion_III}) when the equation is modified to include multispecies effects in a long plasma.  Earlier, Gould and LePointe\cite{goul:92} found shifts proportional to $\omegas$ for a long, single species plasma whose radius is comparable to the wall radius.

\subsubsection{Theoretical Limits on Plasma Temperature Effects}
The plasma temperature could affect our measurements through three mechanisms.  First, like plasma charge effects, it conceivably changes $\omegar$ through changes in the sampled confinement potential anharmonicities; as before, this should not affect the $l=2$ mode.  Second, relativistic effects will change the cyclotron frequency by an amount proportional to the plasma temperature.  For the data in Fig.~\ref{fig:CleanRotScan}, where the temperatures are below $100\,\mathrm{K}$, the resulting shifts are $0.2\,\mathrm{kHz}$ ($0.01\,\mathrm{ppm}$) or lower. Third, there are finite Larmor radius (FLR) affects on the Bernstein-like modes discussed earlier. These effects have been considered by Gould\cite{goul:95} and Dubin.\cite{dubi:13} Gould suggests that these effects scale as $\baromegar(\rL/\rp)$, where $\rL$ is the Larmor radius, which makes an error on the order of $0.016\,\mathrm{kHz}$ ($1\,\mathrm{ppb}$) for our parameters.

\subsubsection{Self-consistency Checks of Plasma Effects}
Within the data in Fig.~\ref{fig:CleanRotScan}, we explored the effect of allowing deviations of the rotation frequency from its calculated value by introducing a frequency deviation parameter $\epsilon$, and fitting with the rotation frequency set to $(1+\epsilon)\omegaz^2/2\omegac$. Such frequency deviations could be evidence of a density effect as the density goes up with increasing bounce frequency, and, hence, rotation frequency.  With this additional fit parameter, we find that $\epsilon=0.007\pm0.008$, consistent with zero, and that the measured cyclotron frequency changes by about $0.07\,\mathrm{kHz}$ ($0.004\,\mathrm{ppm}$).

\subsection{Frequency Pulling from Collisions}
When damped, oscillator frequencies generally diminish. In our specific problem, we would expect an analogous ``pulling'' effect from collisions.  Indeed, using the collision frequency $\nu$ for the inverse decoherence time $\tau$, Eq.~(\ref{eq:DampedEOM}) does exhibit pulling scaling as $0.5(\nu/\omegac)^2\omegar/\omegac$.  This effect is second order compared to the damping itself. Since our collision model is very crude, we will use the pulling formula for a damped harmonic oscillator, $0.5(\nu/\omegac)^2$, as an upper bound.

The plasmas in Fig.~\ref{fig:CleanRotScan} are in the strongly magnetized regime where the collision frequency is suppressed.  Using the formulas developed by Glinsky, et al.,\cite{glin:92}\ we find that the largest collision frequency for this data occurs at $T\approx 50\,\mathrm{K}$, and is $\nu\approx 100\,\mathrm{kHz}$.  The resulting frequency pulling is insignificant ($0.3\,\mathrm{ppt}$) and will be ignored.

\subsection{Frequency Pulling from Cavities}

If the plasma is in a cavity rather than a waveguide, the coupling between plasma and cavity can pull the plasma resonances toward a nearby natural frequency of the cavity. We estimate the scale of this frequency shift by looking at the radiation rate of the plasma in the cavity: $\delta\omega\sim n\xi \FP\gamma V_\text{plasma}$, where
$\xi=\langle|E_-|^2\rangle_\text{plasma}/\max(|E|^2)$,
$\FP=(3/4\pi^2)\lambda^3(Q/V)$ is the Purcell enhancement factor,
$V$ is the mode volume,
$\gamma=e^2\omegac^2/(3\pi\epsilon_0 \me c^3)$ is the (free-space) Larmor radiation rate,
$\lambda$ is the cavity wavelength,
and $Q$ is the cavity quality factor. This formula can be rewritten as
\begin{equation}
\frac{\delta\omega}{\omegac}\sim Q\frac{\omegap^2}{\omegac^2}\frac{V_\text{plasma}}{V_\text{cavity}}\frac{\langle |E_-|^2\rangle_{\text{plasma}}}{\langle |\mathbf{E}|^2\rangle_\text{cavity}},
\end{equation} where $\mathbf{E}$ is the electric field of the cavity mode and $E_-$ is the synchronous part of the cavity field. For an $l=2$ TE mode in a $Q\sim 1000$ cavity with our typical plasma parameters, the fractional shift is $\delta\omega/\omegac\sim 0.1\,\mathrm{ppt}$. For an $l=1$ TE mode, the cavity field mostly overlaps with the plasma leading to a larger, though still insignificant shift of $\delta\omega/\omegac\sim 1\mbox{--}10\,\mathrm{ppb}$.

\section{Fitting a Cyclotron Frequency}
\label{App:Fit}
We used the data in Fig.~\ref{fig:CleanRotScan}, which shows the temperature versus microwave frequency response for multiple bounce frequencies, to make a highly precise measurement of the magnetic field.  Specifically, we used a set of $60$ microwave frequencies $\{F_i\}$ ($i=1,2\ldots 60$) separated by $4\,\mathrm{kHz}$, and $11$ bounce frequencies ($j=1,2\ldots 11$) separated by $2.5\,\mathrm{MHz}$, and measured the temperature response $\{T_{ij}\}$ at each point. For each bounce frequency, we found a set of temperature subpeaks, which we enumerate by $k$, where $k$ ranges from $1$ to $3\mbox{--}9$ depending on how many subpeaks fit in the frequency scan window.

To find the temperature subpeak center frequencies $\{f_{jk}\}$ for a single bounce frequency $j$, we use the Levenberg-Marquardt algorithm to fit a sum of Lorentzian subpeaks to each $\{T_{ij}\}$. The fit function is:
\begin{equation}
\label{eq:Lorentzian}
T_{\text{fit}}(F_i;\{f_{jk}\})=T_0+\sum_{k}\frac{\AL_{jk}}{1+[(f_{jk}-F_i)/\sigmaL_{jk}]^2}.
\end{equation}
In addition to the desired fit parameters $\{f_{jk}\}$, we fit for several nuisance parameters which are hidden in the argument to $T_{\text{fit}}$ (and later in $E_{\mathrm{T}}$): the subpeak amplitudes $\{\AL_{jk}\}$, the subpeak widths $\{\sigmaL_{jk}\}$, and the unheated temperature $T_0$.

We can then make initial guesses for $\{f_{jk}\}$ by assuming that the subpeaks are spaced by $\fr=f_{\mathrm{z}j}/2\fc$ around the frequency of the unmoving subpeak. The initial guesses for $\{f_{jk}\}$ need to be good because the loss function is not a convex function of these parameters. The initial guess for $T_0$, for the $\{\AL_{jk}\}$, and for the $\{\sigmaL_{jk}\}$ are the minimum temperature of the dataset, the difference between the maximum temperature of the dataset and $T_0$, and twice the microwave frequency separation of $4\,\mathrm{kHz}$, respectively. The function $T_{\text{fit}}$ is fit to the data using the Levenberg-Marquardt algorithm, which minimizes the squared error $E_{\mathrm{T}}$:
\begin{equation}
E_{\mathrm{T}}(\{f_{jk}\},j)=\sum_{i}[T_{ij}-T_{\text{fit}}(F_i;\{f_{jk}\})]^2.
\end{equation}
The fit is run for every bounce frequency $j$, resulting in a complete set of subpeak centers $\{f_{jk}\}$.

Next, we label each subpeak using simple heuristics with an $l_{jk}$ value, and fit the following function to the subpeak centers $\{f_{jk}\}$ to find the cyclotron frequency:
\begin{equation}
f_{\text{fit}}(l_{jk},f_{\mathrm{z}j};\fc)=\fc+(l_{jk}-2)\frac{f_{\mathrm{z}j}^2}{2\fc}.
\end{equation}
The bounce frequencies $f_{\mathrm{z}j}$ are known from electrostatic modeling of our trap; thus, this function has only one fit parameter $\fc$. We fit for $\fc$ by minimizing the squared error $E_f$ using the Levenberg Marquardt algorithm,
\begin{equation}
E_f(\fc)=\sum_{j}\sum_{k}[f_{jk}-f_{\text{fit}}(l_{jk},f_{\mathrm{z}j};\fc)]^2/\sigma^2.
\end{equation}
We estimate the error $\sigma$ in our measured subpeak centers by the microwave frequency separation, $\sigma=4\,\mathrm{kHz}$. The statistical error on the fit parameter $\fc$ is estimated by finding how much $\fc$ needs to change to increase $E_f$ by $1/2$.\cite{cous:95}  For the data in Fig.~\ref{fig:CleanRotScan}, this yields a precision of $26\,\mathrm{ppb}$ for $\fc$ and, by extension, for the magnetic field magnitude.

\section{Alternative Techniques}
\label{App:AltTech}
For Penning-Malmberg traps, including those used in antihydrogen research, ECR magnetometry can be employed immediately as these traps generally have electron sources, provisions for close control of electron plasmas, and plasma temperature diagnostics.  It would be difficult to employ non-ECR magnetometry techniques.  For instance, nuclear magnetic resonance (NMR) and Hall probes would have to be inserted into an experiment on a movable, utterly non-magnetic, UHV compatible stick, which would have to extend from room to cryogenic temperatures.   In some current experiments, the stick would have to be as long as $3\,\mathrm{m}$, which, because of space constraints, would require an extensible stick.  Without a complicated load lock, this might require a thermal cycle and vacuum break.  Thus, magnetometry measurements and physics measurements could not be contemporaneous.

Even if these obstacles could be overcome, NMR and Hall probes would require significant development before they could be employed.  Most NMR sample materials are incompatible with the $4\,\mathrm{K}$ trap environment. Exotic sample materials would be required; for example, pressurized helium-3\cite{niki:14} or aluminum powders.\cite{boro:01}  It is not obvious that $1\,\mathrm{ppm}$ accuracy could be reached with the small samples sizes necessary to achieve the required spatial resolution.

Hall effect sensors are quite temperature sensitive, which would be an issue for sensors positioned at the end of a long stick in a large thermal gradient. They have reproducibility issues when thermally cycled, and are not normally more accurate than $\sim 100\,\mathrm{ppm}$; cryogenic, three-dimensional sensors are not readily available and are difficult to calibrate.\cite{sanf:11}

Magnetic fields can be measured to the few $\mathrm{ppb}$ level\cite{smor:18} using highly specialized Penning traps.  These traps employ very high Q superconducting resonators and superconducting amplifiers\cite{naga:16} and would be difficult to adjust for wide ranging fields.  Further, these techniques require precisely harmonic traps in which it would be difficult to freely move the sensing location.

Atomic spectroscopy magnetometry offers a possibly more attractive option.   Neutral atoms suffer from localization issues, but ions could be localized as easily as electrons.  For example, very accurate magnetometry measurements\cite{shig:11,shen:14} have been made with $\mathrm{Be}^+$, and such ions would also be useful for sympathetic cooling of positrons.\cite{jele:03,mads:14}   Once control of the $\mathrm{Be}^+$ ions was established, the major obstacle to spectroscopy-based magnetometry would be achieving a sufficient signal-to-noise ratio; the solid angle available to collect the emitted photons is very small: approximately $10^{-5}\,\mathrm{sr}$ for the ALPHA experiments.

\bibliographystyle{../apsrevNoAuthorLimit}

\begin{thebibliography}{54}
\expandafter\ifx\csname natexlab\endcsname\relax\def\natexlab#1{#1}\fi
\expandafter\ifx\csname bibnamefont\endcsname\relax
  \def\bibnamefont#1{#1}\fi
\expandafter\ifx\csname bibfnamefont\endcsname\relax
  \def\bibfnamefont#1{#1}\fi
\expandafter\ifx\csname citenamefont\endcsname\relax
  \def\citenamefont#1{#1}\fi
\expandafter\ifx\csname url\endcsname\relax
  \def\url#1{\texttt{#1}}\fi
\expandafter\ifx\csname urlprefix\endcsname\relax\def\urlprefix{URL }\fi
\providecommand{\bibinfo}[2]{#2}
\providecommand{\eprint}[2][]{\url{#2}}

\bibitem[{\citenamefont{Malmberg et~al.}(1982)\citenamefont{Malmberg, Driscoll,
  and White}}]{malm:82}
\bibinfo{author}{\bibfnamefont{J.}~\bibnamefont{Malmberg}},
  \bibinfo{author}{\bibfnamefont{C.}~\bibnamefont{Driscoll}}, \bibnamefont{and}
  \bibinfo{author}{\bibfnamefont{W.}~\bibnamefont{White}},
  \bibinfo{journal}{Physica Scripta} \textbf{\bibinfo{volume}{T2B}},
  \bibinfo{pages}{288} (\bibinfo{year}{1982}).

\bibitem[{\citenamefont{Davidson}(1990)}]{davi:90}
\bibinfo{author}{\bibfnamefont{R.}~\bibnamefont{Davidson}},
  \emph{\bibinfo{title}{Physics of Nonneutral Plasmas}}
  (\bibinfo{publisher}{Addison-Wesley}, \bibinfo{address}{Redwood City},
  \bibinfo{year}{1990}).

\bibitem[{\citenamefont{Amole et~al.}(2014)\citenamefont{Amole, Ashkezari,
  Baquero-Ruiz, Bertsche, Butler, Capra, Cesar, Charlton, Deller, Evetts,
  Eriksson, Fajans, Friesen, Fujiwara, Gill, Gutierrez, Hangst, Hardy, Hayden,
  Isaac, Jonsell, Kurchaninov, Little, Madsen, McKenna, Menary, Napoli,
  Olchanski, Olin, Pusa, {C. {\O}. Rasmussen}, Robicheaux, Sarid, Silveira, So,
  Stracka, Tharp, Thompson, van~der Werf, and Wurtele}}]{amol:14b}
\bibinfo{author}{\bibfnamefont{C.}~\bibnamefont{Amole}},
  \bibinfo{author}{\bibfnamefont{M.~D.} \bibnamefont{Ashkezari}},
  \bibinfo{author}{\bibfnamefont{M.}~\bibnamefont{Baquero-Ruiz}},
  \bibinfo{author}{\bibfnamefont{W.}~\bibnamefont{Bertsche}},
  \bibinfo{author}{\bibfnamefont{E.}~\bibnamefont{Butler}},
  \bibinfo{author}{\bibfnamefont{A.}~\bibnamefont{Capra}},
  \bibinfo{author}{\bibfnamefont{C.~L.} \bibnamefont{Cesar}},
  \bibinfo{author}{\bibfnamefont{M.}~\bibnamefont{Charlton}},
  \bibinfo{author}{\bibfnamefont{A.}~\bibnamefont{Deller}},
  \bibinfo{author}{\bibfnamefont{N.}~\bibnamefont{Evetts}},
  \bibinfo{author}{\bibfnamefont{S.}~\bibnamefont{Eriksson}},
  \bibinfo{author}{\bibfnamefont{J.}~\bibnamefont{Fajans}},
  \bibinfo{author}{\bibfnamefont{T.}~\bibnamefont{Friesen}},
  \bibinfo{author}{\bibfnamefont{M.~C.} \bibnamefont{Fujiwara}},
  \bibinfo{author}{\bibfnamefont{D.~R.} \bibnamefont{Gill}},
  \bibinfo{author}{\bibfnamefont{A.}~\bibnamefont{Gutierrez}},
  \bibinfo{author}{\bibfnamefont{J.~S.} \bibnamefont{Hangst}},
  \bibinfo{author}{\bibfnamefont{W.~N.} \bibnamefont{Hardy}},
  \bibinfo{author}{\bibfnamefont{M.~E.} \bibnamefont{Hayden}},
  \bibinfo{author}{\bibfnamefont{C.~A.} \bibnamefont{Isaac}},
  \bibinfo{author}{\bibfnamefont{S.}~\bibnamefont{Jonsell}},
  \bibinfo{author}{\bibfnamefont{L.}~\bibnamefont{Kurchaninov}},
  \bibinfo{author}{\bibfnamefont{A.}~\bibnamefont{Little}},
  \bibinfo{author}{\bibfnamefont{N.}~\bibnamefont{Madsen}},
  \bibinfo{author}{\bibfnamefont{J.~T.~K.} \bibnamefont{McKenna}},
  \bibinfo{author}{\bibfnamefont{S.}~\bibnamefont{Menary}},
  \bibinfo{author}{\bibfnamefont{S.}~\bibnamefont{Napoli}},
  \bibinfo{author}{\bibfnamefont{K.}~\bibnamefont{Olchanski}},
  \bibinfo{author}{\bibfnamefont{A.}~\bibnamefont{Olin}},
  \bibinfo{author}{\bibfnamefont{P.}~\bibnamefont{Pusa}},
  \bibinfo{author}{\bibnamefont{{C. {\O}. Rasmussen}}},
  \bibinfo{author}{\bibfnamefont{F.}~\bibnamefont{Robicheaux}},
  \bibinfo{author}{\bibfnamefont{E.}~\bibnamefont{Sarid}},
  \bibinfo{author}{\bibfnamefont{D.~M.} \bibnamefont{Silveira}},
  \bibinfo{author}{\bibfnamefont{C.}~\bibnamefont{So}},
  \bibinfo{author}{\bibfnamefont{S.}~\bibnamefont{Stracka}},
  \bibinfo{author}{\bibfnamefont{T.}~\bibnamefont{Tharp}},
  \bibinfo{author}{\bibfnamefont{R.~I.} \bibnamefont{Thompson}},
  \bibinfo{author}{\bibfnamefont{D.~P.} \bibnamefont{van~der Werf}},
  \bibnamefont{and} \bibinfo{author}{\bibfnamefont{J.~S.}
  \bibnamefont{Wurtele}} (\bibinfo{collaboration}{ALPHA Collaboration}),
  \bibinfo{journal}{New Journal of Physics} \textbf{\bibinfo{volume}{16}},
  \bibinfo{pages}{013037} (\bibinfo{year}{2014}).

\bibitem[{\citenamefont{Ahmadi et~al.}(2017)\citenamefont{Ahmadi, Alves, Baker,
  Bertsche, Butler, Capra, Carruth, Cesar, Charlton, Cohen, Collister,
  Eriksson, Evans, Evetts, Fajans, Friesen, Fujiwara, Gill, Gutierrez, Hangst,
  Hardy, Hayden, Isaac, Ishida, Johnson, Jones, Jonsell, Kurchaninov, Madsen,
  Mathers, Maxwell, McKenna, Menary, Michan, Momose, Munich, Nolan, Olchanski,
  Olin, Pusa, Rasmussen, Robicheaux, Sacramento, Sameed, Sarid, Silveira,
  Stracka, Stutter, So, Tharp, Thompson, Thompson, van~der Werf, and
  Wurtele}}]{ahma:17b}
\bibinfo{author}{\bibfnamefont{M.}~\bibnamefont{Ahmadi}},
  \bibinfo{author}{\bibfnamefont{B.~X.~R.} \bibnamefont{Alves}},
  \bibinfo{author}{\bibfnamefont{C.~J.} \bibnamefont{Baker}},
  \bibinfo{author}{\bibfnamefont{W.}~\bibnamefont{Bertsche}},
  \bibinfo{author}{\bibfnamefont{E.}~\bibnamefont{Butler}},
  \bibinfo{author}{\bibfnamefont{A.}~\bibnamefont{Capra}},
  \bibinfo{author}{\bibfnamefont{C.}~\bibnamefont{Carruth}},
  \bibinfo{author}{\bibfnamefont{C.~L.} \bibnamefont{Cesar}},
  \bibinfo{author}{\bibfnamefont{M.}~\bibnamefont{Charlton}},
  \bibinfo{author}{\bibfnamefont{S.}~\bibnamefont{Cohen}},
  \bibinfo{author}{\bibfnamefont{R.}~\bibnamefont{Collister}},
  \bibinfo{author}{\bibfnamefont{S.}~\bibnamefont{Eriksson}},
  \bibinfo{author}{\bibfnamefont{A.}~\bibnamefont{Evans}},
  \bibinfo{author}{\bibfnamefont{N.}~\bibnamefont{Evetts}},
  \bibinfo{author}{\bibfnamefont{J.}~\bibnamefont{Fajans}},
  \bibinfo{author}{\bibfnamefont{T.}~\bibnamefont{Friesen}},
  \bibinfo{author}{\bibfnamefont{M.~C.} \bibnamefont{Fujiwara}},
  \bibinfo{author}{\bibfnamefont{D.~R.} \bibnamefont{Gill}},
  \bibinfo{author}{\bibfnamefont{A.}~\bibnamefont{Gutierrez}},
  \bibinfo{author}{\bibfnamefont{J.~S.} \bibnamefont{Hangst}},
  \bibinfo{author}{\bibfnamefont{W.~N.} \bibnamefont{Hardy}},
  \bibinfo{author}{\bibfnamefont{M.~E.} \bibnamefont{Hayden}},
  \bibinfo{author}{\bibfnamefont{C.~A.} \bibnamefont{Isaac}},
  \bibinfo{author}{\bibfnamefont{A.}~\bibnamefont{Ishida}},
  \bibinfo{author}{\bibfnamefont{M.~A.} \bibnamefont{Johnson}},
  \bibinfo{author}{\bibfnamefont{S.~A.} \bibnamefont{Jones}},
  \bibinfo{author}{\bibfnamefont{S.}~\bibnamefont{Jonsell}},
  \bibinfo{author}{\bibfnamefont{L.}~\bibnamefont{Kurchaninov}},
  \bibinfo{author}{\bibfnamefont{N.}~\bibnamefont{Madsen}},
  \bibinfo{author}{\bibfnamefont{M.}~\bibnamefont{Mathers}},
  \bibinfo{author}{\bibfnamefont{D.}~\bibnamefont{Maxwell}},
  \bibinfo{author}{\bibfnamefont{J.~T.~K.} \bibnamefont{McKenna}},
  \bibinfo{author}{\bibfnamefont{S.}~\bibnamefont{Menary}},
  \bibinfo{author}{\bibfnamefont{J.~M.} \bibnamefont{Michan}},
  \bibinfo{author}{\bibfnamefont{T.}~\bibnamefont{Momose}},
  \bibinfo{author}{\bibfnamefont{J.~J.} \bibnamefont{Munich}},
  \bibinfo{author}{\bibfnamefont{P.}~\bibnamefont{Nolan}},
  \bibinfo{author}{\bibfnamefont{K.}~\bibnamefont{Olchanski}},
  \bibinfo{author}{\bibfnamefont{A.}~\bibnamefont{Olin}},
  \bibinfo{author}{\bibfnamefont{P.}~\bibnamefont{Pusa}},
  \bibinfo{author}{\bibfnamefont{C.~{\O}.} \bibnamefont{Rasmussen}},
  \bibinfo{author}{\bibfnamefont{F.}~\bibnamefont{Robicheaux}},
  \bibinfo{author}{\bibfnamefont{R.~L.} \bibnamefont{Sacramento}},
  \bibinfo{author}{\bibfnamefont{M.}~\bibnamefont{Sameed}},
  \bibinfo{author}{\bibfnamefont{E.}~\bibnamefont{Sarid}},
  \bibinfo{author}{\bibfnamefont{D.~M.} \bibnamefont{Silveira}},
  \bibinfo{author}{\bibfnamefont{S.}~\bibnamefont{Stracka}},
  \bibinfo{author}{\bibfnamefont{G.}~\bibnamefont{Stutter}},
  \bibinfo{author}{\bibfnamefont{C.}~\bibnamefont{So}},
  \bibinfo{author}{\bibfnamefont{T.~D.} \bibnamefont{Tharp}},
  \bibinfo{author}{\bibfnamefont{J.~E.} \bibnamefont{Thompson}},
  \bibinfo{author}{\bibfnamefont{R.~I.} \bibnamefont{Thompson}},
  \bibinfo{author}{\bibfnamefont{D.~P.} \bibnamefont{van~der Werf}},
  \bibnamefont{and} \bibinfo{author}{\bibfnamefont{J.~S.}
  \bibnamefont{Wurtele}} (\bibinfo{collaboration}{ALPHA Collaboration}),
  \bibinfo{journal}{Nature} \textbf{\bibinfo{volume}{548}}, \bibinfo{pages}{66}
  (\bibinfo{year}{2017}).

\bibitem[{\citenamefont{Ahmadi et~al.}(2018{\natexlab{a}})\citenamefont{Ahmadi,
  Alves, Baker, Bertsche, Capra, Carruth, Cesar, Charlton, Cohen, Collister,
  Eriksson, Evans, Evetts, Fajans, Friesen, Fujiwara, Gill, Hangst, Hardy,
  Hayden, Isaac, Johnson, Jones, Jones, Jonsell, Khramov, Knapp, Kurchaninov,
  Madsen, Maxwell, McKenna, Menary, Momose, Munich, Olchanski, Olin, Pusa,
  Rasmussen, Robicheaux, Sacramento, Sameed, Sarid, Silveira, Stutter, So,
  Tharp, Thompson, van~der Werf, and Wurtele}}]{ahma:18a}
\bibinfo{author}{\bibfnamefont{M.}~\bibnamefont{Ahmadi}},
  \bibinfo{author}{\bibfnamefont{B.~X.~R.} \bibnamefont{Alves}},
  \bibinfo{author}{\bibfnamefont{C.~J.} \bibnamefont{Baker}},
  \bibinfo{author}{\bibfnamefont{W.}~\bibnamefont{Bertsche}},
  \bibinfo{author}{\bibfnamefont{A.}~\bibnamefont{Capra}},
  \bibinfo{author}{\bibfnamefont{C.}~\bibnamefont{Carruth}},
  \bibinfo{author}{\bibfnamefont{C.~L.} \bibnamefont{Cesar}},
  \bibinfo{author}{\bibfnamefont{M.}~\bibnamefont{Charlton}},
  \bibinfo{author}{\bibfnamefont{S.}~\bibnamefont{Cohen}},
  \bibinfo{author}{\bibfnamefont{R.}~\bibnamefont{Collister}},
  \bibinfo{author}{\bibfnamefont{S.}~\bibnamefont{Eriksson}},
  \bibinfo{author}{\bibfnamefont{A.}~\bibnamefont{Evans}},
  \bibinfo{author}{\bibfnamefont{N.}~\bibnamefont{Evetts}},
  \bibinfo{author}{\bibfnamefont{J.}~\bibnamefont{Fajans}},
  \bibinfo{author}{\bibfnamefont{T.}~\bibnamefont{Friesen}},
  \bibinfo{author}{\bibfnamefont{M.~C.} \bibnamefont{Fujiwara}},
  \bibinfo{author}{\bibfnamefont{D.~R.} \bibnamefont{Gill}},
  \bibinfo{author}{\bibfnamefont{J.~S.} \bibnamefont{Hangst}},
  \bibinfo{author}{\bibfnamefont{W.~N.} \bibnamefont{Hardy}},
  \bibinfo{author}{\bibfnamefont{M.~E.} \bibnamefont{Hayden}},
  \bibinfo{author}{\bibfnamefont{C.~A.} \bibnamefont{Isaac}},
  \bibinfo{author}{\bibfnamefont{M.~A.} \bibnamefont{Johnson}},
  \bibinfo{author}{\bibfnamefont{J.~M.} \bibnamefont{Jones}},
  \bibinfo{author}{\bibfnamefont{S.~A.} \bibnamefont{Jones}},
  \bibinfo{author}{\bibfnamefont{S.}~\bibnamefont{Jonsell}},
  \bibinfo{author}{\bibfnamefont{A.}~\bibnamefont{Khramov}},
  \bibinfo{author}{\bibfnamefont{P.}~\bibnamefont{Knapp}},
  \bibinfo{author}{\bibfnamefont{L.}~\bibnamefont{Kurchaninov}},
  \bibinfo{author}{\bibfnamefont{N.}~\bibnamefont{Madsen}},
  \bibinfo{author}{\bibfnamefont{D.}~\bibnamefont{Maxwell}},
  \bibinfo{author}{\bibfnamefont{J.~T.~K.} \bibnamefont{McKenna}},
  \bibinfo{author}{\bibfnamefont{S.}~\bibnamefont{Menary}},
  \bibinfo{author}{\bibfnamefont{T.}~\bibnamefont{Momose}},
  \bibinfo{author}{\bibfnamefont{J.~J.} \bibnamefont{Munich}},
  \bibinfo{author}{\bibfnamefont{K.}~\bibnamefont{Olchanski}},
  \bibinfo{author}{\bibfnamefont{A.}~\bibnamefont{Olin}},
  \bibinfo{author}{\bibfnamefont{P.}~\bibnamefont{Pusa}},
  \bibinfo{author}{\bibfnamefont{C.~{\O}.} \bibnamefont{Rasmussen}},
  \bibinfo{author}{\bibfnamefont{F.}~\bibnamefont{Robicheaux}},
  \bibinfo{author}{\bibfnamefont{R.~L.} \bibnamefont{Sacramento}},
  \bibinfo{author}{\bibfnamefont{M.}~\bibnamefont{Sameed}},
  \bibinfo{author}{\bibfnamefont{E.}~\bibnamefont{Sarid}},
  \bibinfo{author}{\bibfnamefont{D.~M.} \bibnamefont{Silveira}},
  \bibinfo{author}{\bibfnamefont{G.}~\bibnamefont{Stutter}},
  \bibinfo{author}{\bibfnamefont{C.}~\bibnamefont{So}},
  \bibinfo{author}{\bibfnamefont{T.~D.} \bibnamefont{Tharp}},
  \bibinfo{author}{\bibfnamefont{R.~I.} \bibnamefont{Thompson}},
  \bibinfo{author}{\bibfnamefont{D.~P.} \bibnamefont{van~der Werf}},
  \bibnamefont{and} \bibinfo{author}{\bibfnamefont{J.~S.}
  \bibnamefont{Wurtele}}, \bibinfo{journal}{Nature}
  \textbf{\bibinfo{volume}{557}}, \bibinfo{pages}{71}
  (\bibinfo{year}{2018}{\natexlab{a}}).

\bibitem[{\citenamefont{Ahmadi et~al.}(2018{\natexlab{b}})\citenamefont{Ahmadi,
  Alves, Baker, Bertsche, Capra, Carruth, Cesar, Charlton, Cohen, Collister,
  Eriksson, Evans, Evetts, Fajans, Friesen, Fujiwara, Gill, Hangst, Hardy,
  Hayden, Hunter, Isaac, Johnson, Jones, Jones, Jonsell, Khramov, Knapp,
  Kurchaninov, Madsen, Maxwell, McKenna, Menary, Michan, Momose, Munich,
  Olchanski, Olin, Pusa, Rasmussen, Robicheaux, Sacramento, Sameed, Sarid,
  Silveira, Starko, Stutter, So, Tharp, Thompson, van~der Werf, and
  Wurtele}}]{ahma:18b}
\bibinfo{author}{\bibfnamefont{M.}~\bibnamefont{Ahmadi}},
  \bibinfo{author}{\bibfnamefont{B.~X.~R.} \bibnamefont{Alves}},
  \bibinfo{author}{\bibfnamefont{C.~J.} \bibnamefont{Baker}},
  \bibinfo{author}{\bibfnamefont{W.}~\bibnamefont{Bertsche}},
  \bibinfo{author}{\bibfnamefont{A.}~\bibnamefont{Capra}},
  \bibinfo{author}{\bibfnamefont{C.}~\bibnamefont{Carruth}},
  \bibinfo{author}{\bibfnamefont{C.~L.} \bibnamefont{Cesar}},
  \bibinfo{author}{\bibfnamefont{M.}~\bibnamefont{Charlton}},
  \bibinfo{author}{\bibfnamefont{S.}~\bibnamefont{Cohen}},
  \bibinfo{author}{\bibfnamefont{R.}~\bibnamefont{Collister}},
  \bibinfo{author}{\bibfnamefont{S.}~\bibnamefont{Eriksson}},
  \bibinfo{author}{\bibfnamefont{A.}~\bibnamefont{Evans}},
  \bibinfo{author}{\bibfnamefont{N.}~\bibnamefont{Evetts}},
  \bibinfo{author}{\bibfnamefont{J.}~\bibnamefont{Fajans}},
  \bibinfo{author}{\bibfnamefont{T.}~\bibnamefont{Friesen}},
  \bibinfo{author}{\bibfnamefont{M.~C.} \bibnamefont{Fujiwara}},
  \bibinfo{author}{\bibfnamefont{D.~R.} \bibnamefont{Gill}},
  \bibinfo{author}{\bibfnamefont{J.~S.} \bibnamefont{Hangst}},
  \bibinfo{author}{\bibfnamefont{W.~N.} \bibnamefont{Hardy}},
  \bibinfo{author}{\bibfnamefont{M.~E.} \bibnamefont{Hayden}},
  \bibinfo{author}{\bibfnamefont{E.~D.} \bibnamefont{Hunter}},
  \bibinfo{author}{\bibfnamefont{C.~A.} \bibnamefont{Isaac}},
  \bibinfo{author}{\bibfnamefont{M.~A.} \bibnamefont{Johnson}},
  \bibinfo{author}{\bibfnamefont{J.~M.} \bibnamefont{Jones}},
  \bibinfo{author}{\bibfnamefont{S.~A.} \bibnamefont{Jones}},
  \bibinfo{author}{\bibfnamefont{S.}~\bibnamefont{Jonsell}},
  \bibinfo{author}{\bibfnamefont{A.}~\bibnamefont{Khramov}},
  \bibinfo{author}{\bibfnamefont{P.}~\bibnamefont{Knapp}},
  \bibinfo{author}{\bibfnamefont{L.}~\bibnamefont{Kurchaninov}},
  \bibinfo{author}{\bibfnamefont{N.}~\bibnamefont{Madsen}},
  \bibinfo{author}{\bibfnamefont{D.}~\bibnamefont{Maxwell}},
  \bibinfo{author}{\bibfnamefont{J.~T.~K.} \bibnamefont{McKenna}},
  \bibinfo{author}{\bibfnamefont{S.}~\bibnamefont{Menary}},
  \bibinfo{author}{\bibfnamefont{J.~M.} \bibnamefont{Michan}},
  \bibinfo{author}{\bibfnamefont{T.}~\bibnamefont{Momose}},
  \bibinfo{author}{\bibfnamefont{J.~J.} \bibnamefont{Munich}},
  \bibinfo{author}{\bibfnamefont{K.}~\bibnamefont{Olchanski}},
  \bibinfo{author}{\bibfnamefont{A.}~\bibnamefont{Olin}},
  \bibinfo{author}{\bibfnamefont{P.}~\bibnamefont{Pusa}},
  \bibinfo{author}{\bibfnamefont{C.~{\O}.} \bibnamefont{Rasmussen}},
  \bibinfo{author}{\bibfnamefont{F.}~\bibnamefont{Robicheaux}},
  \bibinfo{author}{\bibfnamefont{R.~L.} \bibnamefont{Sacramento}},
  \bibinfo{author}{\bibfnamefont{M.}~\bibnamefont{Sameed}},
  \bibinfo{author}{\bibfnamefont{E.}~\bibnamefont{Sarid}},
  \bibinfo{author}{\bibfnamefont{D.~M.} \bibnamefont{Silveira}},
  \bibinfo{author}{\bibfnamefont{D.~M.} \bibnamefont{Starko}},
  \bibinfo{author}{\bibfnamefont{G.}~\bibnamefont{Stutter}},
  \bibinfo{author}{\bibfnamefont{C.}~\bibnamefont{So}},
  \bibinfo{author}{\bibfnamefont{T.~D.} \bibnamefont{Tharp}},
  \bibinfo{author}{\bibfnamefont{R.~I.} \bibnamefont{Thompson}},
  \bibinfo{author}{\bibfnamefont{D.~P.} \bibnamefont{van~der Werf}},
  \bibnamefont{and} \bibinfo{author}{\bibfnamefont{J.~S.}
  \bibnamefont{Wurtele}}, \bibinfo{journal}{Nature}
  (\bibinfo{year}{2018}{\natexlab{b}}).

\bibitem[{\citenamefont{Amole et~al.}(2013)\citenamefont{Amole, Ashkezari,
  Baquero-Ruiz, Bertsche, Butler, Capra, Cesar, Charlton, Eriksson, Fajans,
  Friesen, Fujiwara, Gill, Gutierrez, Hangst, Hardy, Hayden, Isaac, Jonsell,
  Kurchaninov, Little, Madsen, McKenna, Menary, Napoli, Nolan, Olin, Pusa,
  Rasmussen, Robicheaux, Sarid, Silveira, So, Thompson, van~der Werf, Wurtele,
  Zhmoginov, and Charman}}]{amol:13}
\bibinfo{author}{\bibfnamefont{C.}~\bibnamefont{Amole}},
  \bibinfo{author}{\bibfnamefont{M.~D.} \bibnamefont{Ashkezari}},
  \bibinfo{author}{\bibfnamefont{M.}~\bibnamefont{Baquero-Ruiz}},
  \bibinfo{author}{\bibfnamefont{W.}~\bibnamefont{Bertsche}},
  \bibinfo{author}{\bibfnamefont{E.}~\bibnamefont{Butler}},
  \bibinfo{author}{\bibfnamefont{A.}~\bibnamefont{Capra}},
  \bibinfo{author}{\bibfnamefont{C.~L.} \bibnamefont{Cesar}},
  \bibinfo{author}{\bibfnamefont{M.}~\bibnamefont{Charlton}},
  \bibinfo{author}{\bibfnamefont{S.}~\bibnamefont{Eriksson}},
  \bibinfo{author}{\bibfnamefont{J.}~\bibnamefont{Fajans}},
  \bibinfo{author}{\bibfnamefont{T.}~\bibnamefont{Friesen}},
  \bibinfo{author}{\bibfnamefont{M.~C.} \bibnamefont{Fujiwara}},
  \bibinfo{author}{\bibfnamefont{D.~R.} \bibnamefont{Gill}},
  \bibinfo{author}{\bibfnamefont{A.}~\bibnamefont{Gutierrez}},
  \bibinfo{author}{\bibfnamefont{J.~S.} \bibnamefont{Hangst}},
  \bibinfo{author}{\bibfnamefont{W.~N.} \bibnamefont{Hardy}},
  \bibinfo{author}{\bibfnamefont{M.~E.} \bibnamefont{Hayden}},
  \bibinfo{author}{\bibfnamefont{C.~A.} \bibnamefont{Isaac}},
  \bibinfo{author}{\bibfnamefont{S.}~\bibnamefont{Jonsell}},
  \bibinfo{author}{\bibfnamefont{L.}~\bibnamefont{Kurchaninov}},
  \bibinfo{author}{\bibfnamefont{A.}~\bibnamefont{Little}},
  \bibinfo{author}{\bibfnamefont{N.}~\bibnamefont{Madsen}},
  \bibinfo{author}{\bibfnamefont{J.}~\bibnamefont{McKenna}},
  \bibinfo{author}{\bibfnamefont{S.}~\bibnamefont{Menary}},
  \bibinfo{author}{\bibfnamefont{S.~C.} \bibnamefont{Napoli}},
  \bibinfo{author}{\bibfnamefont{P.}~\bibnamefont{Nolan}},
  \bibinfo{author}{\bibfnamefont{A.}~\bibnamefont{Olin}},
  \bibinfo{author}{\bibfnamefont{P.}~\bibnamefont{Pusa}},
  \bibinfo{author}{\bibfnamefont{C.~{\O}.} \bibnamefont{Rasmussen}},
  \bibinfo{author}{\bibfnamefont{F.}~\bibnamefont{Robicheaux}},
  \bibinfo{author}{\bibfnamefont{E.}~\bibnamefont{Sarid}},
  \bibinfo{author}{\bibfnamefont{D.~M.} \bibnamefont{Silveira}},
  \bibinfo{author}{\bibfnamefont{C.}~\bibnamefont{So}},
  \bibinfo{author}{\bibfnamefont{R.~I.} \bibnamefont{Thompson}},
  \bibinfo{author}{\bibfnamefont{D.}~\bibnamefont{van~der Werf}},
  \bibinfo{author}{\bibfnamefont{J.~S.} \bibnamefont{Wurtele}},
  \bibinfo{author}{\bibfnamefont{A.~I.} \bibnamefont{Zhmoginov}},
  \bibnamefont{and} \bibinfo{author}{\bibfnamefont{A.~E.}
  \bibnamefont{Charman}} (\bibinfo{collaboration}{ALPHA Collaboration}),
  \bibinfo{journal}{Nat. Commun.} \textbf{\bibinfo{volume}{4}},
  \bibinfo{pages}{1785} (\bibinfo{year}{2013}).

\bibitem[{\citenamefont{Zhmoginov et~al.}(2013)\citenamefont{Zhmoginov,
  Charman, Shalloo, Fajans, and Wurtele}}]{zhmo:13}
\bibinfo{author}{\bibfnamefont{A.~I.} \bibnamefont{Zhmoginov}},
  \bibinfo{author}{\bibfnamefont{A.~E.} \bibnamefont{Charman}},
  \bibinfo{author}{\bibfnamefont{R.}~\bibnamefont{Shalloo}},
  \bibinfo{author}{\bibfnamefont{J.}~\bibnamefont{Fajans}}, \bibnamefont{and}
  \bibinfo{author}{\bibfnamefont{J.~S.} \bibnamefont{Wurtele}},
  \bibinfo{journal}{Class. and Quantum Grav.} \textbf{\bibinfo{volume}{30}},
  \bibinfo{pages}{205014} (\bibinfo{year}{2013}).

\bibitem[{\citenamefont{Hamilton et~al.}(2014)\citenamefont{Hamilton,
  Zhmoginov, Robicheaux, Fajans, Wurtele, and M\"uller}}]{hami:14}
\bibinfo{author}{\bibfnamefont{P.}~\bibnamefont{Hamilton}},
  \bibinfo{author}{\bibfnamefont{A.}~\bibnamefont{Zhmoginov}},
  \bibinfo{author}{\bibfnamefont{F.}~\bibnamefont{Robicheaux}},
  \bibinfo{author}{\bibfnamefont{J.}~\bibnamefont{Fajans}},
  \bibinfo{author}{\bibfnamefont{J.~S.} \bibnamefont{Wurtele}},
  \bibnamefont{and} \bibinfo{author}{\bibfnamefont{H.}~\bibnamefont{M\"uller}},
  \bibinfo{journal}{Phys. Rev. Lett.} \textbf{\bibinfo{volume}{112}},
  \bibinfo{pages}{121102} (\bibinfo{year}{2014}).

\bibitem[{\citenamefont{O'Neil}(1980{\natexlab{a}})}]{onei:80a}
\bibinfo{author}{\bibfnamefont{T.~M.} \bibnamefont{O'Neil}},
  \bibinfo{journal}{Phys.\ Fluids} \textbf{\bibinfo{volume}{23}},
  \bibinfo{pages}{725} (\bibinfo{year}{1980}{\natexlab{a}}).

\bibitem[{\citenamefont{Gould}(1995)}]{goul:95}
\bibinfo{author}{\bibfnamefont{R.~W.} \bibnamefont{Gould}},
  \bibinfo{journal}{Physics of Plasmas} \textbf{\bibinfo{volume}{2}},
  \bibinfo{pages}{1404} (\bibinfo{year}{1995}).

\bibitem[{\citenamefont{Dubin}(2013)}]{dubi:13}
\bibinfo{author}{\bibfnamefont{D.~H.~E.} \bibnamefont{Dubin}},
  \bibinfo{journal}{Physics of Plasmas} \textbf{\bibinfo{volume}{20}},
  \bibinfo{pages}{042120} (\bibinfo{year}{2013}).

\bibitem[{\citenamefont{Gould and LaPointe}(1992)}]{goul:92}
\bibinfo{author}{\bibfnamefont{R.~W.} \bibnamefont{Gould}} \bibnamefont{and}
  \bibinfo{author}{\bibfnamefont{M.~A.} \bibnamefont{LaPointe}},
  \bibinfo{journal}{Physics of Fluids B: Plasma Physics}
  \textbf{\bibinfo{volume}{4}}, \bibinfo{pages}{2038} (\bibinfo{year}{1992}).

\bibitem[{\citenamefont{Sarid et~al.}(1995)\citenamefont{Sarid, Anderegg, and
  Driscoll}}]{sari:95}
\bibinfo{author}{\bibfnamefont{E.}~\bibnamefont{Sarid}},
  \bibinfo{author}{\bibfnamefont{F.}~\bibnamefont{Anderegg}}, \bibnamefont{and}
  \bibinfo{author}{\bibfnamefont{C.~F.} \bibnamefont{Driscoll}},
  \bibinfo{journal}{Physics of Plasmas} \textbf{\bibinfo{volume}{2}},
  \bibinfo{pages}{2895} (\bibinfo{year}{1995}).

\bibitem[{\citenamefont{Affolter et~al.}(2015)\citenamefont{Affolter, Anderegg,
  Dubin, and Driscoll}}]{affo:15}
\bibinfo{author}{\bibfnamefont{M.}~\bibnamefont{Affolter}},
  \bibinfo{author}{\bibfnamefont{F.}~\bibnamefont{Anderegg}},
  \bibinfo{author}{\bibfnamefont{D.~H.~E.} \bibnamefont{Dubin}},
  \bibnamefont{and} \bibinfo{author}{\bibfnamefont{C.~F.}
  \bibnamefont{Driscoll}}, \bibinfo{journal}{Physics of Plasmas}
  \textbf{\bibinfo{volume}{22}}, \bibinfo{pages}{055701}
  (\bibinfo{year}{2015}).

\bibitem[{\citenamefont{Peurrung and Fajans}(1993)}]{peur:93a}
\bibinfo{author}{\bibfnamefont{A.~J.} \bibnamefont{Peurrung}} \bibnamefont{and}
  \bibinfo{author}{\bibfnamefont{J.}~\bibnamefont{Fajans}},
  \bibinfo{journal}{Rev.\ Sci.\ Instrum.} \textbf{\bibinfo{volume}{64}},
  \bibinfo{pages}{52} (\bibinfo{year}{1993}).

\bibitem[{\citenamefont{Evans}(2016)}]{evan:16a}
\bibinfo{author}{\bibfnamefont{L.~T.} \bibnamefont{Evans}}, Ph.D. thesis,
  \bibinfo{school}{University of California, Berkeley} (\bibinfo{year}{2016}).

\bibitem[{\citenamefont{Hunter et~al.}(2020)\citenamefont{Hunter, Fajans,
  Lewis, Povilus, Sierra, So, and Zimmer}}]{hunt:20a}
\bibinfo{author}{\bibfnamefont{E.~D.} \bibnamefont{Hunter}},
  \bibinfo{author}{\bibfnamefont{J.}~\bibnamefont{Fajans}},
  \bibinfo{author}{\bibfnamefont{N.~A.} \bibnamefont{Lewis}},
  \bibinfo{author}{\bibfnamefont{A.~P.} \bibnamefont{Povilus}},
  \bibinfo{author}{\bibfnamefont{C.}~\bibnamefont{Sierra}},
  \bibinfo{author}{\bibfnamefont{C.}~\bibnamefont{So}}, \bibnamefont{and}
  \bibinfo{author}{\bibfnamefont{D.}~\bibnamefont{Zimmer}},
  \emph{\bibinfo{title}{Plasma temperature measurement with a silicon
  photomultiplier ({S}i{P}{M})}} (\bibinfo{year}{2020}), \bibinfo{note}{in
  preparation}.

\bibitem[{\citenamefont{Eggleston et~al.}(1992)\citenamefont{Eggleston,
  Driscoll, Beck, Hyatt, and Malmberg}}]{eggl:92}
\bibinfo{author}{\bibfnamefont{D.~L.} \bibnamefont{Eggleston}},
  \bibinfo{author}{\bibfnamefont{C.~F.} \bibnamefont{Driscoll}},
  \bibinfo{author}{\bibfnamefont{B.~R.} \bibnamefont{Beck}},
  \bibinfo{author}{\bibfnamefont{A.~W.} \bibnamefont{Hyatt}}, \bibnamefont{and}
  \bibinfo{author}{\bibfnamefont{J.~H.} \bibnamefont{Malmberg}},
  \bibinfo{journal}{Phys.\ Fluids B} \textbf{\bibinfo{volume}{4}},
  \bibinfo{pages}{3432} (\bibinfo{year}{1992}).

\bibitem[{\citenamefont{Friesen}(2014)}]{frie:14}
\bibinfo{author}{\bibfnamefont{T.~P.} \bibnamefont{Friesen}}, Ph.D. thesis,
  \bibinfo{school}{University of Calgary} (\bibinfo{year}{2014}),
  \urlprefix\url{http://alpha.web.cern.ch/sites/alpha.web.cern.ch/files/FriesenThesis.pdf}.

\bibitem[{\citenamefont{Dubin}(1991)}]{dubi:91}
\bibinfo{author}{\bibfnamefont{D.~H.~E.} \bibnamefont{Dubin}},
  \bibinfo{journal}{Phys. Rev. Lett.} \textbf{\bibinfo{volume}{66}},
  \bibinfo{pages}{2076} (\bibinfo{year}{1991}).

\bibitem[{\citenamefont{Tinkle et~al.}(1994)\citenamefont{Tinkle, Greaves.,
  Surko, Spencer, and Mason}}]{tink:94}
\bibinfo{author}{\bibfnamefont{M.~D.} \bibnamefont{Tinkle}},
  \bibinfo{author}{\bibfnamefont{R.~G.} \bibnamefont{Greaves.}},
  \bibinfo{author}{\bibfnamefont{C.~M.} \bibnamefont{Surko}},
  \bibinfo{author}{\bibfnamefont{R.~L.} \bibnamefont{Spencer}},
  \bibnamefont{and} \bibinfo{author}{\bibfnamefont{G.~W.} \bibnamefont{Mason}},
  \bibinfo{journal}{Phys. Rev. Lett.} \textbf{\bibinfo{volume}{72}},
  \bibinfo{pages}{352} (\bibinfo{year}{1994}).

\bibitem[{\citenamefont{Danielson et~al.}(2007)\citenamefont{Danielson, Weber,
  and Surko}}]{dani:07}
\bibinfo{author}{\bibfnamefont{J.~R.} \bibnamefont{Danielson}},
  \bibinfo{author}{\bibfnamefont{T.~R.} \bibnamefont{Weber}}, \bibnamefont{and}
  \bibinfo{author}{\bibfnamefont{C.~M.} \bibnamefont{Surko}},
  \bibinfo{journal}{Applied Physics Letters} \textbf{\bibinfo{volume}{90}},
  \bibinfo{pages}{081503} (\bibinfo{year}{2007}).

\bibitem[{\citenamefont{{Smorra} et~al.}(2015)\citenamefont{{Smorra}, {Mooser},
  {Franke}, {Nagahama}, {Schneider}, {Higuchi}, {Gorp}, {Blaum}, {Matsuda},
  {Quint}, {Walz}, {Yamazaki}, and {Ulmer}}}]{smor:15}
\bibinfo{author}{\bibfnamefont{C.}~\bibnamefont{{Smorra}}},
  \bibinfo{author}{\bibfnamefont{A.}~\bibnamefont{{Mooser}}},
  \bibinfo{author}{\bibfnamefont{K.}~\bibnamefont{{Franke}}},
  \bibinfo{author}{\bibfnamefont{H.}~\bibnamefont{{Nagahama}}},
  \bibinfo{author}{\bibfnamefont{G.}~\bibnamefont{{Schneider}}},
  \bibinfo{author}{\bibfnamefont{T.}~\bibnamefont{{Higuchi}}},
  \bibinfo{author}{\bibfnamefont{S.~V.} \bibnamefont{{Gorp}}},
  \bibinfo{author}{\bibfnamefont{K.}~\bibnamefont{{Blaum}}},
  \bibinfo{author}{\bibfnamefont{Y.}~\bibnamefont{{Matsuda}}},
  \bibinfo{author}{\bibfnamefont{W.}~\bibnamefont{{Quint}}},
  \bibinfo{author}{\bibfnamefont{J.}~\bibnamefont{{Walz}}},
  \bibinfo{author}{\bibfnamefont{Y.}~\bibnamefont{{Yamazaki}}},
  \bibnamefont{and} \bibinfo{author}{\bibfnamefont{S.}~\bibnamefont{{Ulmer}}},
  \bibinfo{journal}{International Journal of Mass Spectrometry}
  \textbf{\bibinfo{volume}{389}}, \bibinfo{pages}{10} (\bibinfo{year}{2015}).

\bibitem[{\citenamefont{Ahmadi et~al.}(2018{\natexlab{c}})\citenamefont{Ahmadi,
  Alves, Baker, Bertsche, Capra, Carruth, Cesar, Charlton, Cohen, Collister,
  Eriksson, Evans, Evetts, Fajans, Friesen, Fujiwara, Gill, Hangst, Hardy,
  Hayden, Isaac, Johnson, Jones, Jonsell, Kurchaninov, Madsen, Mathers,
  Maxwell, McKenna, Menary, Momose, Munich, Olchanski, Olin, Pusa, Rasmussen,
  Robicheaux, Sacramento, Sameed, Sarid, Silveira, So, Stutter, Tharp,
  Thompson, Thompson, van~der Werf, and Wurtele}}]{ahma:18}
\bibinfo{author}{\bibfnamefont{M.}~\bibnamefont{Ahmadi}},
  \bibinfo{author}{\bibfnamefont{B.~X.~R.} \bibnamefont{Alves}},
  \bibinfo{author}{\bibfnamefont{C.~J.} \bibnamefont{Baker}},
  \bibinfo{author}{\bibfnamefont{W.}~\bibnamefont{Bertsche}},
  \bibinfo{author}{\bibfnamefont{A.}~\bibnamefont{Capra}},
  \bibinfo{author}{\bibfnamefont{C.}~\bibnamefont{Carruth}},
  \bibinfo{author}{\bibfnamefont{C.~L.} \bibnamefont{Cesar}},
  \bibinfo{author}{\bibfnamefont{M.}~\bibnamefont{Charlton}},
  \bibinfo{author}{\bibfnamefont{S.}~\bibnamefont{Cohen}},
  \bibinfo{author}{\bibfnamefont{R.}~\bibnamefont{Collister}},
  \bibinfo{author}{\bibfnamefont{S.}~\bibnamefont{Eriksson}},
  \bibinfo{author}{\bibfnamefont{A.}~\bibnamefont{Evans}},
  \bibinfo{author}{\bibfnamefont{N.}~\bibnamefont{Evetts}},
  \bibinfo{author}{\bibfnamefont{J.}~\bibnamefont{Fajans}},
  \bibinfo{author}{\bibfnamefont{T.}~\bibnamefont{Friesen}},
  \bibinfo{author}{\bibfnamefont{M.~C.} \bibnamefont{Fujiwara}},
  \bibinfo{author}{\bibfnamefont{D.~R.} \bibnamefont{Gill}},
  \bibinfo{author}{\bibfnamefont{J.~S.} \bibnamefont{Hangst}},
  \bibinfo{author}{\bibfnamefont{W.~N.} \bibnamefont{Hardy}},
  \bibinfo{author}{\bibfnamefont{M.~E.} \bibnamefont{Hayden}},
  \bibinfo{author}{\bibfnamefont{C.~A.} \bibnamefont{Isaac}},
  \bibinfo{author}{\bibfnamefont{M.~A.} \bibnamefont{Johnson}},
  \bibinfo{author}{\bibfnamefont{S.~A.} \bibnamefont{Jones}},
  \bibinfo{author}{\bibfnamefont{S.}~\bibnamefont{Jonsell}},
  \bibinfo{author}{\bibfnamefont{L.}~\bibnamefont{Kurchaninov}},
  \bibinfo{author}{\bibfnamefont{N.}~\bibnamefont{Madsen}},
  \bibinfo{author}{\bibfnamefont{M.}~\bibnamefont{Mathers}},
  \bibinfo{author}{\bibfnamefont{D.}~\bibnamefont{Maxwell}},
  \bibinfo{author}{\bibfnamefont{J.~T.~K.} \bibnamefont{McKenna}},
  \bibinfo{author}{\bibfnamefont{S.}~\bibnamefont{Menary}},
  \bibinfo{author}{\bibfnamefont{T.}~\bibnamefont{Momose}},
  \bibinfo{author}{\bibfnamefont{J.~J.} \bibnamefont{Munich}},
  \bibinfo{author}{\bibfnamefont{K.}~\bibnamefont{Olchanski}},
  \bibinfo{author}{\bibfnamefont{A.}~\bibnamefont{Olin}},
  \bibinfo{author}{\bibfnamefont{P.}~\bibnamefont{Pusa}},
  \bibinfo{author}{\bibfnamefont{C.~O.} \bibnamefont{Rasmussen}},
  \bibinfo{author}{\bibfnamefont{F.}~\bibnamefont{Robicheaux}},
  \bibinfo{author}{\bibfnamefont{R.~L.} \bibnamefont{Sacramento}},
  \bibinfo{author}{\bibfnamefont{M.}~\bibnamefont{Sameed}},
  \bibinfo{author}{\bibfnamefont{E.}~\bibnamefont{Sarid}},
  \bibinfo{author}{\bibfnamefont{D.~M.} \bibnamefont{Silveira}},
  \bibinfo{author}{\bibfnamefont{C.}~\bibnamefont{So}},
  \bibinfo{author}{\bibfnamefont{G.}~\bibnamefont{Stutter}},
  \bibinfo{author}{\bibfnamefont{T.~D.} \bibnamefont{Tharp}},
  \bibinfo{author}{\bibfnamefont{J.~E.} \bibnamefont{Thompson}},
  \bibinfo{author}{\bibfnamefont{R.~I.} \bibnamefont{Thompson}},
  \bibinfo{author}{\bibfnamefont{D.~P.} \bibnamefont{van~der Werf}},
  \bibnamefont{and} \bibinfo{author}{\bibfnamefont{J.~S.}
  \bibnamefont{Wurtele}} (\bibinfo{collaboration}{ALPHA Collaboration}),
  \bibinfo{journal}{Phys. Rev. Lett.} \textbf{\bibinfo{volume}{120}},
  \bibinfo{pages}{025001} (\bibinfo{year}{2018}{\natexlab{c}}).

\bibitem[{\citenamefont{Danielson and Surko}(2005)}]{dani:05}
\bibinfo{author}{\bibfnamefont{J.~R.} \bibnamefont{Danielson}}
  \bibnamefont{and} \bibinfo{author}{\bibfnamefont{C.~M.} \bibnamefont{Surko}},
  \bibinfo{journal}{Phys.\ Rev.\ Lett.} \textbf{\bibinfo{volume}{94}},
  \bibinfo{pages}{035001} (\bibinfo{year}{2005}).

\bibitem[{\citenamefont{Andresen et~al.}(2010)\citenamefont{Andresen,
  Ashkezari, Baquero-Ruiz, Bertsche, Bowe, Butler, Cesar, Chapman, Charlton,
  Fajans, Friesen, Fujiwara, Gill, Hangst, Hardy, Hayano, Hayden, Humphries,
  Hydomako, Jonsell, Kurchaninov, Lambo, Madsen, Menary, Nolan, Olchanski,
  Olin, Povilus, Pusa, Robicheaux, Sarid, Silveira, So, Storey, Thompson,
  van~der Werf, Wilding, Wurtele, and Yamazaki}}]{andr:10}
\bibinfo{author}{\bibfnamefont{G.~B.} \bibnamefont{Andresen}},
  \bibinfo{author}{\bibfnamefont{M.~D.} \bibnamefont{Ashkezari}},
  \bibinfo{author}{\bibfnamefont{M.}~\bibnamefont{Baquero-Ruiz}},
  \bibinfo{author}{\bibfnamefont{W.}~\bibnamefont{Bertsche}},
  \bibinfo{author}{\bibfnamefont{P.~D.} \bibnamefont{Bowe}},
  \bibinfo{author}{\bibfnamefont{E.}~\bibnamefont{Butler}},
  \bibinfo{author}{\bibfnamefont{C.~L.} \bibnamefont{Cesar}},
  \bibinfo{author}{\bibfnamefont{S.}~\bibnamefont{Chapman}},
  \bibinfo{author}{\bibfnamefont{M.}~\bibnamefont{Charlton}},
  \bibinfo{author}{\bibfnamefont{J.}~\bibnamefont{Fajans}},
  \bibinfo{author}{\bibfnamefont{T.}~\bibnamefont{Friesen}},
  \bibinfo{author}{\bibfnamefont{M.~C.} \bibnamefont{Fujiwara}},
  \bibinfo{author}{\bibfnamefont{D.~R.} \bibnamefont{Gill}},
  \bibinfo{author}{\bibfnamefont{J.~S.} \bibnamefont{Hangst}},
  \bibinfo{author}{\bibfnamefont{W.~N.} \bibnamefont{Hardy}},
  \bibinfo{author}{\bibfnamefont{R.~S.} \bibnamefont{Hayano}},
  \bibinfo{author}{\bibfnamefont{M.~E.} \bibnamefont{Hayden}},
  \bibinfo{author}{\bibfnamefont{A.}~\bibnamefont{Humphries}},
  \bibinfo{author}{\bibfnamefont{R.}~\bibnamefont{Hydomako}},
  \bibinfo{author}{\bibfnamefont{S.}~\bibnamefont{Jonsell}},
  \bibinfo{author}{\bibfnamefont{L.}~\bibnamefont{Kurchaninov}},
  \bibinfo{author}{\bibfnamefont{R.}~\bibnamefont{Lambo}},
  \bibinfo{author}{\bibfnamefont{N.}~\bibnamefont{Madsen}},
  \bibinfo{author}{\bibfnamefont{S.}~\bibnamefont{Menary}},
  \bibinfo{author}{\bibfnamefont{P.}~\bibnamefont{Nolan}},
  \bibinfo{author}{\bibfnamefont{K.}~\bibnamefont{Olchanski}},
  \bibinfo{author}{\bibfnamefont{A.}~\bibnamefont{Olin}},
  \bibinfo{author}{\bibfnamefont{A.}~\bibnamefont{Povilus}},
  \bibinfo{author}{\bibfnamefont{P.}~\bibnamefont{Pusa}},
  \bibinfo{author}{\bibfnamefont{F.}~\bibnamefont{Robicheaux}},
  \bibinfo{author}{\bibfnamefont{E.}~\bibnamefont{Sarid}},
  \bibinfo{author}{\bibfnamefont{D.~M.} \bibnamefont{Silveira}},
  \bibinfo{author}{\bibfnamefont{C.}~\bibnamefont{So}},
  \bibinfo{author}{\bibfnamefont{J.~W.} \bibnamefont{Storey}},
  \bibinfo{author}{\bibfnamefont{R.~I.} \bibnamefont{Thompson}},
  \bibinfo{author}{\bibfnamefont{D.~P.} \bibnamefont{van~der Werf}},
  \bibinfo{author}{\bibfnamefont{D.}~\bibnamefont{Wilding}},
  \bibinfo{author}{\bibfnamefont{J.~S.} \bibnamefont{Wurtele}},
  \bibnamefont{and} \bibinfo{author}{\bibfnamefont{Y.}~\bibnamefont{Yamazaki}}
  (\bibinfo{collaboration}{ALPHA Collaboration}), \bibinfo{journal}{Phys. Rev.
  Lett.} \textbf{\bibinfo{volume}{105}}, \bibinfo{pages}{013003}
  (\bibinfo{year}{2010}).

\bibitem[{\citenamefont{Prasad and O'Neil}(1979)}]{pras:79}
\bibinfo{author}{\bibfnamefont{S.~A.} \bibnamefont{Prasad}} \bibnamefont{and}
  \bibinfo{author}{\bibfnamefont{T.~M.} \bibnamefont{O'Neil}},
  \bibinfo{journal}{Phys.\ Fluids} \textbf{\bibinfo{volume}{22}},
  \bibinfo{pages}{278} (\bibinfo{year}{1979}).

\bibitem[{\citenamefont{Peurrung and Fajans}(1990)}]{peur:90}
\bibinfo{author}{\bibfnamefont{A.~J.} \bibnamefont{Peurrung}} \bibnamefont{and}
  \bibinfo{author}{\bibfnamefont{J.}~\bibnamefont{Fajans}},
  \bibinfo{journal}{Phys.\ Fluids B} \textbf{\bibinfo{volume}{2}},
  \bibinfo{pages}{693} (\bibinfo{year}{1990}).

\bibitem[{\citenamefont{Spencer et~al.}(1993)\citenamefont{Spencer, Rasband,
  and Vanfleet}}]{spen:93}
\bibinfo{author}{\bibfnamefont{R.~L.} \bibnamefont{Spencer}},
  \bibinfo{author}{\bibfnamefont{S.~N.} \bibnamefont{Rasband}},
  \bibnamefont{and} \bibinfo{author}{\bibfnamefont{R.~R.}
  \bibnamefont{Vanfleet}}, \bibinfo{journal}{Physics of Fluids B: Plasma
  Physics} \textbf{\bibinfo{volume}{5}}, \bibinfo{pages}{4267}
  (\bibinfo{year}{1993}).

\bibitem[{\citenamefont{Byrne and Farago}(1965)}]{byrn:65}
\bibinfo{author}{\bibfnamefont{J.}~\bibnamefont{Byrne}} \bibnamefont{and}
  \bibinfo{author}{\bibfnamefont{P.~S.} \bibnamefont{Farago}},
  \bibinfo{journal}{Proceedings of the Physical Society}
  \textbf{\bibinfo{volume}{86}}, \bibinfo{pages}{801} (\bibinfo{year}{1965}).

\bibitem[{\citenamefont{Jeffries et~al.}(1983)\citenamefont{Jeffries, Barlow,
  and Dunn}}]{jeff:83}
\bibinfo{author}{\bibfnamefont{J.}~\bibnamefont{Jeffries}},
  \bibinfo{author}{\bibfnamefont{S.}~\bibnamefont{Barlow}}, \bibnamefont{and}
  \bibinfo{author}{\bibfnamefont{G.}~\bibnamefont{Dunn}},
  \bibinfo{journal}{International Journal of Mass Spectrometry and Ion
  Processes} \textbf{\bibinfo{volume}{54}}, \bibinfo{pages}{169 }
  (\bibinfo{year}{1983}), ISSN \bibinfo{issn}{0168-1176}.

\bibitem[{com()}]{coms}
\emph{\bibinfo{title}{Comsol multiphysics v. 5.4}},
  \bibinfo{howpublished}{www.comsol.com/documentation}.

\bibitem[{\citenamefont{Hunter et~al.}(2018)\citenamefont{Hunter, Evetts,
  Fajans, Hardy, Landsberger, Mcpeters, and Wurtele}}]{hunt:18}
\bibinfo{author}{\bibfnamefont{E.~D.} \bibnamefont{Hunter}},
  \bibinfo{author}{\bibfnamefont{N.}~\bibnamefont{Evetts}},
  \bibinfo{author}{\bibfnamefont{J.}~\bibnamefont{Fajans}},
  \bibinfo{author}{\bibfnamefont{W.~N.} \bibnamefont{Hardy}},
  \bibinfo{author}{\bibfnamefont{H.}~\bibnamefont{Landsberger}},
  \bibinfo{author}{\bibfnamefont{R.}~\bibnamefont{Mcpeters}}, \bibnamefont{and}
  \bibinfo{author}{\bibfnamefont{J.~S.} \bibnamefont{Wurtele}},
  \bibinfo{journal}{Physics of Plasmas} \textbf{\bibinfo{volume}{25}},
  \bibinfo{pages}{011602} (\bibinfo{year}{2018}).

\bibitem[{\citenamefont{O'Neil}(1983)}]{onei:83}
\bibinfo{author}{\bibfnamefont{T.~M.} \bibnamefont{O'Neil}},
  \bibinfo{journal}{Phys.\ Fluids} \textbf{\bibinfo{volume}{26}},
  \bibinfo{pages}{2128} (\bibinfo{year}{1983}).

\bibitem[{\citenamefont{Savory et~al.}(2011)\citenamefont{Savory, Kaiser,
  McKenna, Xian, Blakney, Rodgers, Hendrickson, and Marshall}}]{savo:11}
\bibinfo{author}{\bibfnamefont{J.~J.} \bibnamefont{Savory}},
  \bibinfo{author}{\bibfnamefont{N.~K.} \bibnamefont{Kaiser}},
  \bibinfo{author}{\bibfnamefont{A.~M.} \bibnamefont{McKenna}},
  \bibinfo{author}{\bibfnamefont{F.}~\bibnamefont{Xian}},
  \bibinfo{author}{\bibfnamefont{G.~T.} \bibnamefont{Blakney}},
  \bibinfo{author}{\bibfnamefont{R.~P.} \bibnamefont{Rodgers}},
  \bibinfo{author}{\bibfnamefont{C.~L.} \bibnamefont{Hendrickson}},
  \bibnamefont{and} \bibinfo{author}{\bibfnamefont{A.~G.}
  \bibnamefont{Marshall}}, \bibinfo{journal}{Analytical Chemistry}
  \textbf{\bibinfo{volume}{83}}, \bibinfo{pages}{1732} (\bibinfo{year}{2011}).

\bibitem[{\citenamefont{Smorra et~al.}(2018)\citenamefont{Smorra, Blessing,
  Borchert, Devlin, Harrington, Higuchi, Morgner, Nagahama, Sellner, Bohman,
  Mooser, Schneider, Sch\"{o}n, Wiesinger, Blaum, Matsuda, Ospelkaus, Quint,
  Walz, Yamazaki, and Ulmer}}]{smor:18}
\bibinfo{author}{\bibfnamefont{C.}~\bibnamefont{Smorra}},
  \bibinfo{author}{\bibfnamefont{P.~E.} \bibnamefont{Blessing}},
  \bibinfo{author}{\bibfnamefont{M.~J.} \bibnamefont{Borchert}},
  \bibinfo{author}{\bibfnamefont{J.~A.} \bibnamefont{Devlin}},
  \bibinfo{author}{\bibfnamefont{J.~A.} \bibnamefont{Harrington}},
  \bibinfo{author}{\bibfnamefont{T.}~\bibnamefont{Higuchi}},
  \bibinfo{author}{\bibfnamefont{J.}~\bibnamefont{Morgner}},
  \bibinfo{author}{\bibfnamefont{H.}~\bibnamefont{Nagahama}},
  \bibinfo{author}{\bibfnamefont{S.}~\bibnamefont{Sellner}},
  \bibinfo{author}{\bibfnamefont{M.~A.} \bibnamefont{Bohman}},
  \bibinfo{author}{\bibfnamefont{A.~H.} \bibnamefont{Mooser}},
  \bibinfo{author}{\bibfnamefont{G.~L.} \bibnamefont{Schneider}},
  \bibinfo{author}{\bibfnamefont{N.}~\bibnamefont{Sch\"{o}n}},
  \bibinfo{author}{\bibfnamefont{M.}~\bibnamefont{Wiesinger}},
  \bibinfo{author}{\bibfnamefont{K.}~\bibnamefont{Blaum}},
  \bibinfo{author}{\bibfnamefont{Y.}~\bibnamefont{Matsuda}},
  \bibinfo{author}{\bibfnamefont{C.}~\bibnamefont{Ospelkaus}},
  \bibinfo{author}{\bibfnamefont{W.}~\bibnamefont{Quint}},
  \bibinfo{author}{\bibfnamefont{J.}~\bibnamefont{Walz}},
  \bibinfo{author}{\bibfnamefont{Y.}~\bibnamefont{Yamazaki}}, \bibnamefont{and}
  \bibinfo{author}{\bibfnamefont{S.}~\bibnamefont{Ulmer}},
  \bibinfo{journal}{Hyperfine Interactions} \textbf{\bibinfo{volume}{239}}
  (\bibinfo{year}{2018}).

\bibitem[{\citenamefont{O'Neil}(1980{\natexlab{b}})}]{onei:80}
\bibinfo{author}{\bibfnamefont{T.~M.} \bibnamefont{O'Neil}},
  \bibinfo{journal}{Phys.\ Fluids} \textbf{\bibinfo{volume}{23}},
  \bibinfo{pages}{2216} (\bibinfo{year}{1980}{\natexlab{b}}).

\bibitem[{\citenamefont{Fajans}(2003)}]{faja:03}
\bibinfo{author}{\bibfnamefont{J.}~\bibnamefont{Fajans}},
  \bibinfo{journal}{Phys.\ Plasmas} \textbf{\bibinfo{volume}{10}},
  \bibinfo{pages}{1209} (\bibinfo{year}{2003}).

\bibitem[{\citenamefont{Dubin and Tsidulko}(2011)}]{dubi:11}
\bibinfo{author}{\bibfnamefont{D.~H.~E.} \bibnamefont{Dubin}} \bibnamefont{and}
  \bibinfo{author}{\bibfnamefont{Y.~A.} \bibnamefont{Tsidulko}},
  \bibinfo{journal}{Physics of Plasmas} \textbf{\bibinfo{volume}{18}},
  \bibinfo{pages}{062114} (\bibinfo{year}{2011}).

\bibitem[{\citenamefont{Kabantsev et~al.}(2014)\citenamefont{Kabantsev,
  Driscoll, Dubin, and Tsidulko}}]{kaba:14}
\bibinfo{author}{\bibfnamefont{A.~A.} \bibnamefont{Kabantsev}},
  \bibinfo{author}{\bibfnamefont{C.~F.} \bibnamefont{Driscoll}},
  \bibinfo{author}{\bibfnamefont{D.~H.~E.} \bibnamefont{Dubin}},
  \bibnamefont{and} \bibinfo{author}{\bibfnamefont{Y.~A.}
  \bibnamefont{Tsidulko}}, \bibinfo{journal}{Journal of Plasma Physics}
  \textbf{\bibinfo{volume}{81}} (\bibinfo{year}{2014}).

\bibitem[{\citenamefont{Dubin}(2017)}]{dubi:17}
\bibinfo{author}{\bibfnamefont{D.~H.~E.} \bibnamefont{Dubin}},
  \bibinfo{journal}{Physics of Plasmas} \textbf{\bibinfo{volume}{24}},
  \bibinfo{pages}{112120} (\bibinfo{year}{2017}).

\bibitem[{\citenamefont{V\"ollinger}(2003)}]{voll:03}
\bibinfo{author}{\bibfnamefont{C.}~\bibnamefont{V\"ollinger}}, Ph.D. thesis,
  \bibinfo{school}{Technical University of Berlin} (\bibinfo{year}{2003}).

\bibitem[{\citenamefont{Wineland and Dehmelt}(1975)}]{wine:75a}
\bibinfo{author}{\bibfnamefont{D.}~\bibnamefont{Wineland}} \bibnamefont{and}
  \bibinfo{author}{\bibfnamefont{H.}~\bibnamefont{Dehmelt}},
  \bibinfo{journal}{International Journal of Mass Spectrometry and Ion Physics}
  \textbf{\bibinfo{volume}{16}}, \bibinfo{pages}{338} (\bibinfo{year}{1975}).

\bibitem[{\citenamefont{Glinsky et~al.}(1992)\citenamefont{Glinsky, O'Neil,
  Rosenbluth, Tsuruta, and Ichimaru}}]{glin:92}
\bibinfo{author}{\bibfnamefont{M.~E.} \bibnamefont{Glinsky}},
  \bibinfo{author}{\bibfnamefont{T.~M.} \bibnamefont{O'Neil}},
  \bibinfo{author}{\bibfnamefont{M.~N.} \bibnamefont{Rosenbluth}},
  \bibinfo{author}{\bibfnamefont{K.}~\bibnamefont{Tsuruta}}, \bibnamefont{and}
  \bibinfo{author}{\bibfnamefont{S.}~\bibnamefont{Ichimaru}},
  \bibinfo{journal}{Phys.\ Fluids B} \textbf{\bibinfo{volume}{4}},
  \bibinfo{pages}{1156} (\bibinfo{year}{1992}).

\bibitem[{\citenamefont{Cousins}(1995)}]{cous:95}
\bibinfo{author}{\bibfnamefont{R.~D.} \bibnamefont{Cousins}},
  \bibinfo{journal}{American Journal of Physics} \textbf{\bibinfo{volume}{63}},
  \bibinfo{pages}{398} (\bibinfo{year}{1995}).

\bibitem[{\citenamefont{Nikiel et~al.}(2014)\citenamefont{Nikiel, Bl\"{u}mler,
  Heil, Hehn, Karpuk, Maul, Otten, Schreiber, and Terekhov}}]{niki:14}
\bibinfo{author}{\bibfnamefont{A.}~\bibnamefont{Nikiel}},
  \bibinfo{author}{\bibfnamefont{P.}~\bibnamefont{Bl\"{u}mler}},
  \bibinfo{author}{\bibfnamefont{W.}~\bibnamefont{Heil}},
  \bibinfo{author}{\bibfnamefont{M.}~\bibnamefont{Hehn}},
  \bibinfo{author}{\bibfnamefont{S.}~\bibnamefont{Karpuk}},
  \bibinfo{author}{\bibfnamefont{A.}~\bibnamefont{Maul}},
  \bibinfo{author}{\bibfnamefont{E.}~\bibnamefont{Otten}},
  \bibinfo{author}{\bibfnamefont{L.~M.} \bibnamefont{Schreiber}},
  \bibnamefont{and} \bibinfo{author}{\bibfnamefont{M.}~\bibnamefont{Terekhov}},
  \bibinfo{journal}{The European Physical Journal D}
  \textbf{\bibinfo{volume}{68}} (\bibinfo{year}{2014}).

\bibitem[{\citenamefont{Borovikov et~al.}(2001)\citenamefont{Borovikov,
  Fedurin, Karpov, Korshunov, Kuper, Kuzin, Mamkin, Medvedko, Mezentsev,
  Repkov, Shkaruba, Shubin, and Veremeenko}}]{boro:01}
\bibinfo{author}{\bibfnamefont{V.}~\bibnamefont{Borovikov}},
  \bibinfo{author}{\bibfnamefont{M.}~\bibnamefont{Fedurin}},
  \bibinfo{author}{\bibfnamefont{G.}~\bibnamefont{Karpov}},
  \bibinfo{author}{\bibfnamefont{D.}~\bibnamefont{Korshunov}},
  \bibinfo{author}{\bibfnamefont{E.}~\bibnamefont{Kuper}},
  \bibinfo{author}{\bibfnamefont{M.}~\bibnamefont{Kuzin}},
  \bibinfo{author}{\bibfnamefont{V.}~\bibnamefont{Mamkin}},
  \bibinfo{author}{\bibfnamefont{A.}~\bibnamefont{Medvedko}},
  \bibinfo{author}{\bibfnamefont{N.}~\bibnamefont{Mezentsev}},
  \bibinfo{author}{\bibfnamefont{V.}~\bibnamefont{Repkov}},
  \bibinfo{author}{\bibfnamefont{V.}~\bibnamefont{Shkaruba}},
  \bibinfo{author}{\bibfnamefont{E.}~\bibnamefont{Shubin}}, \bibnamefont{and}
  \bibinfo{author}{\bibfnamefont{V.}~\bibnamefont{Veremeenko}},
  \bibinfo{journal}{Nuclear Instruments and Methods in Physics Research Section
  A: Accelerators, Spectrometers, Detectors and Associated Equipment}
  \textbf{\bibinfo{volume}{467-468}}, \bibinfo{pages}{198}
  (\bibinfo{year}{2001}).

\bibitem[{\citenamefont{Sanfilippo}(2011)}]{sanf:11}
\bibinfo{author}{\bibfnamefont{S.}~\bibnamefont{Sanfilippo}}
  (\bibinfo{year}{2011}), \eprint{http://arxiv.org/abs/1103.1271v1}.

\bibitem[{\citenamefont{Nagahama et~al.}(2016)\citenamefont{Nagahama,
  Schneider, Mooser, Smorra, Sellner, Harrington, Higuchi, Borchert, Tanaka,
  Besirli, Blaum, Matsuda, Ospelkaus, Quint, Walz, Yamazaki, and
  Ulmer}}]{naga:16}
\bibinfo{author}{\bibfnamefont{H.}~\bibnamefont{Nagahama}},
  \bibinfo{author}{\bibfnamefont{G.}~\bibnamefont{Schneider}},
  \bibinfo{author}{\bibfnamefont{A.}~\bibnamefont{Mooser}},
  \bibinfo{author}{\bibfnamefont{C.}~\bibnamefont{Smorra}},
  \bibinfo{author}{\bibfnamefont{S.}~\bibnamefont{Sellner}},
  \bibinfo{author}{\bibfnamefont{J.}~\bibnamefont{Harrington}},
  \bibinfo{author}{\bibfnamefont{T.}~\bibnamefont{Higuchi}},
  \bibinfo{author}{\bibfnamefont{M.}~\bibnamefont{Borchert}},
  \bibinfo{author}{\bibfnamefont{T.}~\bibnamefont{Tanaka}},
  \bibinfo{author}{\bibfnamefont{M.}~\bibnamefont{Besirli}},
  \bibinfo{author}{\bibfnamefont{K.}~\bibnamefont{Blaum}},
  \bibinfo{author}{\bibfnamefont{Y.}~\bibnamefont{Matsuda}},
  \bibinfo{author}{\bibfnamefont{C.}~\bibnamefont{Ospelkaus}},
  \bibinfo{author}{\bibfnamefont{W.}~\bibnamefont{Quint}},
  \bibinfo{author}{\bibfnamefont{J.}~\bibnamefont{Walz}},
  \bibinfo{author}{\bibfnamefont{Y.}~\bibnamefont{Yamazaki}}, \bibnamefont{and}
  \bibinfo{author}{\bibfnamefont{S.}~\bibnamefont{Ulmer}},
  \bibinfo{journal}{Review of Scientific Instruments}
  \textbf{\bibinfo{volume}{87}}, \bibinfo{pages}{113305}
  (\bibinfo{year}{2016}).

\bibitem[{\citenamefont{Shiga et~al.}(2011)\citenamefont{Shiga, Itano, and
  Bollinger}}]{shig:11}
\bibinfo{author}{\bibfnamefont{N.}~\bibnamefont{Shiga}},
  \bibinfo{author}{\bibfnamefont{W.~M.} \bibnamefont{Itano}}, \bibnamefont{and}
  \bibinfo{author}{\bibfnamefont{J.~J.} \bibnamefont{Bollinger}},
  \bibinfo{journal}{Physical Review A} \textbf{\bibinfo{volume}{84}}
  (\bibinfo{year}{2011}).

\bibitem[{\citenamefont{Shen et~al.}(2014)\citenamefont{Shen, Borodin, and
  Schiller}}]{shen:14}
\bibinfo{author}{\bibfnamefont{J.}~\bibnamefont{Shen}},
  \bibinfo{author}{\bibfnamefont{A.}~\bibnamefont{Borodin}}, \bibnamefont{and}
  \bibinfo{author}{\bibfnamefont{S.}~\bibnamefont{Schiller}},
  \bibinfo{journal}{The European Physical Journal D}
  \textbf{\bibinfo{volume}{68}} (\bibinfo{year}{2014}).

\bibitem[{\citenamefont{Jelenkovi\'{c}
  et~al.}(2003)\citenamefont{Jelenkovi\'{c}, Newbury, Bollinger, Itano, and
  Mitchell}}]{jele:03}
\bibinfo{author}{\bibfnamefont{B.~M.} \bibnamefont{Jelenkovi\'{c}}},
  \bibinfo{author}{\bibfnamefont{A.~S.} \bibnamefont{Newbury}},
  \bibinfo{author}{\bibfnamefont{J.~J.} \bibnamefont{Bollinger}},
  \bibinfo{author}{\bibfnamefont{W.~M.} \bibnamefont{Itano}}, \bibnamefont{and}
  \bibinfo{author}{\bibfnamefont{T.~B.} \bibnamefont{Mitchell}},
  \bibinfo{journal}{Phys. Rev. A} \textbf{\bibinfo{volume}{67}},
  \bibinfo{pages}{063406} (\bibinfo{year}{2003}).

\bibitem[{\citenamefont{Madsen et~al.}(2014)\citenamefont{Madsen, Robicheaux,
  and Jonsell}}]{mads:14}
\bibinfo{author}{\bibfnamefont{N.}~\bibnamefont{Madsen}},
  \bibinfo{author}{\bibfnamefont{F.}~\bibnamefont{Robicheaux}},
  \bibnamefont{and} \bibinfo{author}{\bibfnamefont{S.}~\bibnamefont{Jonsell}},
  \bibinfo{journal}{New J. Phys.} \textbf{\bibinfo{volume}{16}},
  \bibinfo{pages}{063046} (\bibinfo{year}{2014}).

\end{thebibliography}

\end{document}